# Weak Segregation Theory and Non-Conventional Morphologies in the Ternary ABC Triblock Copolymers.

*Igor Ya. Erukhimovich*

Moscow State University, Moscow 119899 Russia

ierukhs@polly.phys.msu.ru

Non-Conventional Morphologies in ABC Copolymers.

## ABSTRACT

The statistical theory of microphase separation in the ternary ABC triblock copolymers is presented and the corresponding phase diagrams are built both for the linear and miktoarm copolymers. For this purpose the Leibler weak segregation theory in molten diblock copolymers is generalized to multi-component monodisperse block copolymers with due regard for the 2$^{nd}$ shell harmonics contributions defined in the paper. The Hildebrand approximation for the χ-parameters is used. The physical meaning of this and alternative choices for the χ-parameters is discussed. The symmetric linear $A_f B_{1-2f} C_f$ copolymers with the middle block non-selective with respect to the side ones are shown to undergo the continuous ODT not only into the lamellar phase but also, instead, into various non-conventional cubic phases (depending on the middle block composition it could be the simple cubic, face-centered cubic or non-centrosymmetric phase, which reveals the symmetry of $I4_132$ space group No.214 first predicted to appear in molten block copolymers). For asymmetric linear ABC copolymers a region of compositions is found where the weakly segregated gyroid (double gyroid) phase exists between the planar hexagonal and lamellar or one of the non-conventional cubic phases up to the very critical point. In contrast, the miktoarm (star) ABC block copolymers with one of its arm non-selective with respect to the two others are shown to reveal a pronounced tendency towards strong segregation, which is preceded by increase of stability of the conventional BCC phase and a peculiar weakly segregated BCC phase (BCC$_3$), where the dominant harmonics belong to the 3$^{rd}$ coordination sphere of the reciprocal lattice. The validity region of the developed theory is discussed and outlined in the composition triangles both for linear and miktoarm copolymers. We present also the list of the 2$^{nd}$ shell harmonics (SAXS reflections) allowed and prohibited in some of the non-conventional morphologies due to the weak segregation considerations and comparison of our results with the preceding SCFT treatment of the ABC copolymers by Matsen.

## 1. Introduction

One of the most interesting phenomena occurring in solutions and melts of copolymers is so-called microphase separation or order-disorder transition (ODT), which is known[1-3] to be due to the instability of the uniform state with respect to spatial fluctuations of the polymer concentration having a finite period $L$ and wave number $q^*=2\pi/L$. With changing (typically decreasing) the temperature, the ODT is followed by some order-order transitions.[4-8] The various morphologies emerging as a result of these transitions have been attracted much interest due to both interesting physics underlying their formation and numerous possible technological applications (in particular, as photonic crystals[9]).

These morphologies are characterized by certain spatially inhomogeneous profiles of the densities of the repeated polymer units (monomers), which have a crystal symmetry, the size of the corresponding unit cell being typically of the order of magnitude of macromolecular gyration radius. In the case the amplitude of such density modulations is small a common theoretical framework for these transitions is provided by the weak segregation (WS) theory.[2] (In the physical literature it is called the weak crystallization theory.[10-13]) As consistent with the mean field approximation of the WS theory, which was first developed by Leibler[2] for molten diblock copolymers, there is the following typical succession of the 1st order phase transitions occurring with decrease of temperature: the uniform (disordered) phase (DIS) - body-centered cubic lattice (BCC) - hexagonal planar lattice (HEX) - lamellar structure (L). (We refer further to the phases BCC, HEX and LAM as the conventional ones and to all other phases as non-conventional.) The same sequence of the transitions was found theoretically to hold for molten diblock copolymers within the strong segregation limit[14] and, within the weak segregation limit, for molten triblock and trigraft[15,16,19] as well as polyblock, polygraft,[17,19] and star[18,19] binary AB copolymers.

For some special values of parameters characterizing the structure of the aforementioned systems all the phase transition lines calculated in the mean field approximation of the WS theory merge at certain temperature (the set of these values and temperatures is referred to as that of the critical points), where a continuous (the 2nd order) phase transition from the disordered to lamellar phase was predicted to occur. For $A_{fN}B_{(1-f)N}$ diblock copolymers such a parameter is just the composition $f$ and the critical point is observed for the symmetric diblock copolymer $f_A = f_B = f_{crit} = 0.5$ (if the repeated units of both blocks have the same Kuhn lengths $a$ and excluded volumes $v$). Some other phases like simple cubic (SC) and face-centered cubic (FCC) could exist as metastable. For some special models[12] these phases were shown also to become thermodynamically stable.

As first observed experimentally in refs 20,21 there is one more equilibrium phase in molten $AB$ diblock and star copolymers in an intermediate segregation regime. This phase, which is also commonly encountered in lipid-water and surfactant systems,[22,23] is characterized by $Ia\bar{3}d$ space group symmetry and referred to as the (double) gyroid (G). The bi-continuous morphology characteristic of this phase has been attracted much interest during the last decade and a few of theoretical papers has been published to describe its properties. In particular, an efficient numerical procedure was presented[24,25] to calculate the mean-field phase diagrams (including the G phase) for the aforementioned copolymer systems in an intermediate segregation regime. A specific feature of these phase diagrams[24,25] is existence of two triple points $f_A = f_1^t < f_{crit}$, $T = T_1^t$ and $f_A = f_2^t > f_{crit}$, $T = T_2^t$, where three phases HEX, G and LAM coexist. Therewith, the conventional sequence DIS-BCC-HEX-LAM and non-conventional one DIS-BCC-HEX-G-LAM hold for compositions within and out of the interval $(f_1^t, f_2^t)$, respectively. It is worth to notice here that the experimental[21] and SCFT[24] phase diagrams are in a qualitative agreement only. The quantitative discrepancy is due to the fact that so called fluctuation corrections[26] neglected both within the SCFT and mean field WS theory are far from being minor as first shown for block copolymers within the WS theory by Fredrickson and Helfand[27] (see also refs 12, 28-32).

Even more spectacular new morphologies could be observed in the molten ternary ABC linear[33-39] and star (miktoarm)[40,41] block copolymers (see also refs 8,42 and references therein). The corresponding phase diagrams were calculated by Stadler et al.[37] and by Matsen[43] using the Alexander-de Gennes and SCFT approaches, respectively. Within the WS theory (both with and without the fluctuation corrections) no phase diagrams of these systems were calculated by that time. Moreover, Matsen[43] has queried about the very usefulness of the WS theory for these systems: "In the diblock system, the spinodal line is a good indicator of the order-disorder transition (ODT), but we doubt that is the case in the triblock system since the single harmonic approximation is no longer reliable. We attribute their unusual behavior … to the failure of the single-harmonic approximation. In general, it will be necessary to retain numerous harmonics even at weak segregations, and this will make Leibler-type calculations very involved. Considering that such calculations are only applicable in a narrow region along the ODT, they are surely not worth the effort unless fluctuation corrections are of interest."

It is to show that, on the contrary, the properly used WS theory stays an efficient tool to predict and analyze the non-conventional phase diagrams, which is the purpose of the present paper. In fact, the SCFT itself could provide incorrect results when it is not based on the prior WS analysis. In particular, in the ternary ABC block copolymers we find some new stable non-conventional morphologies overlooked by Matsen[43] as well as very important architecture effects. The subsequent presentation is organized as follows.

In section 2 we introduce three non-conventional cubic phase ($BCC_2$ or single gyroid, $BCC_3$ and $G_2$) and discuss in detail how to describe them within the weak segregation theory. Here we also summarize the results of our general analysis of the non-conventional phase stability, which gives an idea of how some non-conventional morphologies could occur within the first harmonics approximation. It is this analysis that led us to conclude that the existence of triple point(s) different from the critical one is *not* a necessary feature of the phase diagrams of the systems revealing the thermodynamically stable bicontinuous gyroid morphology. This conclusion[44] was the starting point of our study aimed to find, basing on microscopic calculations, those real systems, which would reveal the stable morphologies G, SC, FCC (as well as $BCC_2$ and $G_2$ defined below). As shown in our preliminary communications,[45,46] these systems are some special ternary linear ABC block copolymers. Therefore, in section 3 we present the basics of the weak segregation theory for the multi-component system. The central point of this section is the distinction between the weakly and strongly fluctuating order parameters, which is used to define unambiguously the critical points with respect to the ODT in these systems and build, in a vicinity of this point, a one-parametric (scalar) mean-field rep-resentation of the free energy of the systems with vector order parameter. The higher harmonics' contribution is also naturally (even though partially) included into this scheme. The readers interested mainly in the results may skip section 3, whose purpose is to introduce the one-order and 2nd shell harmonics approximations, validate them near the critical point(s) and, thus, provide the theoreticians with all the basic facts they could need to use our approach themselves. In section 4 we study the spinodal and critical surfaces of molten ABC linear and miktoarm copolymers for different sets of the interaction Flory-Huggins parameters. In section 5 we present an additional discussion of the 2nd shell harmonics approximation validity and build the phase portraits and diagrams for some classes of molten ternary ABC copolymers (both linear and miktoarm). Here we show that the phase behavior of the ABC systems is extremely sensitive to their architecture and discuss which choice of the interaction parameters is physically validated. In section 6 we discuss how the SAXS and SANS data could be used to identify the non-conventional cubic phases. At last, in section 7 we summarize our results.

The fluctuation effects for the non-conventional morphologies will be addressed elsewhere (in the vein of the Brazovskii-Fredrickson-Helfand approximation[12,26-32]).

## 2. General analysis of the non-conventional phases stability in the one-order-parameter systems under weak segregation.

In order to analyze the general conditions ensuring that the non-conventional phases do exist we consider a general class of systems characterized by a local scalar order parameter $\Phi(\mathbf{r})$ and undergoing weak crystallization close to the corresponding critical point. (In this section no explicit peculiarities of the copolymer systems are involved.) It is natural to write down the free energy of such systems as a Landau expansion in powers of $\Phi$ up to the 4th order:

$$\Delta F = (1/2)\int \Phi(\mathbf{r})\left[\tau + \left(q_*^2 + \partial^2/\partial \mathbf{r}^2\right)^2\right]\Phi(\mathbf{r})d\mathbf{r} + \Delta F_3 + \Delta F_4, \qquad (2.1)$$

$$\Delta F_n = (1/n!)\int \Gamma_n(\mathbf{r}_1 - \mathbf{r},..\mathbf{r}_n - \mathbf{r})d\mathbf{r}\prod_{i=1}^{n}\Phi(\mathbf{r}_i)d\mathbf{r}_i, \qquad (2.1a)$$

which in the Fourier-representation reads

$$\Delta F = \int \frac{\Gamma_2(q)}{2} \frac{|\Phi_\mathbf{q}|^2 d\mathbf{q}}{(2\pi)^3} + \Delta F_3 + \Delta F_4, \tag{2.2}$$

$$\Delta F_n = \frac{1}{n!}\int \delta\left(\sum_{i=1}^n \mathbf{q}_i\right) \Gamma_n(\mathbf{q_1},..,\mathbf{q}_n) \prod_{i=1}^n \frac{\Phi(\mathbf{q}_i)d\mathbf{q}_i}{(2\pi)^3}. \tag{2.2a}$$

Hereafter, we refer to functions $f(r)$ and their Fourier transforms $f(\mathbf{q}) = \int d\mathbf{r}\, f(\mathbf{r})\exp(i\mathbf{qr})$ as the same functions in r- and q-representation, respectively, and distinguish them only by the choice of the letters used to denote their arguments; due to the context this convention is not expected to cause any misunderstandings.

The functions $\Gamma_n$ appearing in the cubic and quadric free energy contributions (2.1a) depend on the structure of the system, the function $\Gamma_2(q) = \tau + (q^2 - q_*^2)$ has a minimum at $q = q_*$, which is the generic property of the systems undergoing weak crystallization, and $\tau$ is an effective dimensionless temperature measured from the instability point.

The system morphology is described, within the mean-field approximation, by the order parameter $\Phi(\mathbf{r})$ profile providing the minimum of this free energy. If the symmetry of a morphology is that of a spatial lattice $\mathfrak{R}$, $\Phi(\mathbf{r})$ can be expanded in an infinite series in the Fourier harmonics corresponding to the set of the points of the lattice $\mathfrak{R}^{-1}$ conjugated to $\mathfrak{R}$:

$$\phi(\mathbf{r}) = \sum_{\mathbf{q}_i \in \mathfrak{R}^{-1}} A(\mathbf{q}_i)\exp i(\mathbf{q_i r} + \varphi_i) \tag{2.3}$$

But close to the critical point one takes account[2,13] only of the main harmonics, i.e. of those $2k$ vectors $\mathbf{q}_i$ whose absolute values equal $q_*$:

$$\phi(\mathbf{r}) = A\Psi(\mathbf{r}), \quad \Psi(\mathbf{r}) = \sum_{|\mathbf{q}_i|=q_*} \exp i(\mathbf{q_i r} + \varphi(\mathbf{q}_i)). \tag{2.3a}$$

Note that any physically meaningful function $\phi(\mathbf{r})$ is real and, therefore, $\varphi(\mathbf{q}_i) = -\varphi(-\mathbf{q}_i)$.

It is the choice of the general trial function (2.3) in the form (2.3a), which is called the weak crystallization (or the weak segregation) approximation. Usually (but not necessary, as shown below) the main harmonics belong to the 1st coordination sphere of the chosen conjugated lattice. Therewith, the choice of the phases $\varphi_i$ in the definition (2.2a) of the basic function $\Psi(r)$ is relevant.

Indeed, substituting the trial function (2.3a) into the free energy (2.2) for a morphology $\Re$ with $2k$ main harmonics and minimizing the result with respect to $A$ gives[2]

$$\Delta F_\Re = \min(\tau A_0^2 + \alpha_\Re A_0^3 + \beta_\Re A_0^4) = \frac{\left(3|\alpha_\Re| + \sqrt{9\alpha_\Re^2 - 32\tau\beta_\Re}\right)^3 \left(|\alpha_\Re| - \sqrt{9\alpha_\Re^2 - 32\tau\beta_\Re}\right)}{2^{12}\beta_\Re^3} \quad (2.4)$$

Here we introduced the reduced order parameter amplitude $A_0 = Ak^{1/2}$, cubic vertex

$$\alpha_\Re = \gamma(1)C_\Re, \quad C_\Re = \sum_3^\Re \cos\Omega_j^{(3)} / k^{3/2} \quad (2.5)$$

and quadric vertex

$$\beta_\Re = \frac{\lambda_0(0)}{4k} + \frac{k\sum\lambda_0(h_i) + \sum_4^\Re \lambda(h_1, h_2)\cos\Omega_j^{(4)}}{k^2}. \quad (2.6)$$

In (2.5), (2.6) we use the designations and parameters of Leibler[2]

$$\gamma(h) = \Gamma_3(\mathbf{q}_1, \mathbf{q}_2, \mathbf{q}_3), \quad q_1^2 = q_2^2 = q_*^2, \quad q_3^2 = (\mathbf{q}_1 + \mathbf{q}_2)^2 = hq_*^2, \quad (2.7)$$

$$h_1 = (\mathbf{q_1}+\mathbf{q_2})^2/q_*^2, \quad h_2 = (\mathbf{q_1}+\mathbf{q_3})^2/q_*^2, \quad h_3 = 4 - h_1 - h_2 = (\mathbf{q_1}+\mathbf{q_4})^2/q_*^2, \quad (2.8)$$

$$\lambda(h_1, h_2) = \Gamma_4(\mathbf{q}_1, \mathbf{q}_2, \mathbf{q}_3, \mathbf{q}_4), \quad |\mathbf{q}_i| = q_*, \quad i = 1,..4 \quad (2.9)$$

$$\lambda_0(h) = \Gamma_4(\mathbf{q}, -\mathbf{q}, \mathbf{p}, -\mathbf{p}) = \lambda(0, h), \quad h = (\mathbf{q}+\mathbf{p})^2/q_*^2. \quad (2.10)$$

The phases $\Omega_j^{(3)}, \Omega_j^{(4)}$ are the algebraic sums of the phases $\varphi$ for the triplets and non-coplanar quartets of the vectors $\{\mathbf{q}_i\}$ involved in the definition of $\gamma$ and $\lambda$, respectively, the symbol $\sum_n^\Re$ designates summation over all sets of such $n$ vectors for given morphology $\Re$ (we explain this procedure in more detail below). The first summation in (2.6) is, in fact, over all pairs of non-collinear vectors $\mathbf{q}_i$ and $\mathbf{q}_j$.

In order to demonstrate the effect of the phase shifts $\varphi$, let us consider the systems with the same main harmonics bur different phases.

**The BCC family.** Let the set $\{\mathbf{q}_i\}$ consist of 12 vectors whose relative directions are given, e.g., by the six vectors listed below and the same vectors taken with the opposite sign:

$$\begin{aligned}\mathbf{q}_1 &= (q_*/\sqrt{2})(0,1,-1), & \mathbf{q}_2 &= (q_*/\sqrt{2})(-1,0,1,), & \mathbf{q}_3 &= (q_*/\sqrt{2})(1,-1,0), \\ \mathbf{q}_\mathrm{I} &= (q_*/\sqrt{2})(0,-1,-1), & \mathbf{q}_\mathrm{II} &= (q_*/\sqrt{2})(-1,0,-1,), & \mathbf{q}_\mathrm{III} &= (q_*/\sqrt{2})(-1,-1,0)\end{aligned} \quad (2.11)$$

It follows from (2.11) that the following equalities hold:

$$\mathbf{q}_I = \mathbf{q}_{II} + \mathbf{q}_3, \quad \mathbf{q}_{II} = \mathbf{q}_{III} + \mathbf{q}_1, \quad \mathbf{q}_{III} = \mathbf{q}_I + \mathbf{q}_2, \quad \mathbf{q}_1 + \mathbf{q}_2 + \mathbf{q}_3 = 0 \quad (2.11a)$$

The vectors (2.11) could be visualized as the edges of octahedron[2] or tetrahedron[12] but for our purposes it is convenient to use their planar mapping shown in Figure 1. The arrow circuits in Figure 1 correspond to the equalities (2.11a).

For the conventional BCC, all the phases φ appearing in the definition (2.2a) of the basic function Ψ (and, thus, Ω) are zero. Then

$$\Psi_{BCC}(\mathbf{r})/2 = \cos(x+y) + \cos(z+y) + \cos(x+z) + \cos(x-y) + \cos(z-y) + \cos(x-z) \\ = 2(\cos x \cos y + \cos x \cos z + \cos z \cos y) \quad (2.12)$$

$$\alpha_{BCC} = 8\gamma/6^{3/2}, \quad \beta_{BCC} = [\lambda_0(0) + 8\lambda_0(1) + 2\lambda_0(2) + 4\lambda(1,2,1)]/24. \quad (2.13)$$

Now, if the phases corresponding to three vectors $\mathbf{q}_1$, $\mathbf{q}_2$ and $\mathbf{q}_3$, which form a base of the tetrahedron, and those of three non-coplanar vectors $\mathbf{q}_I$, $\mathbf{q}_{II}$ and $\mathbf{q}_{III}$ (the thin and solid lines, respectively, in Figure 1) equal π/2 and zero, respectively, we get a lattice first discussed in ref 12 and called there the BCC$_2$. For this lattice $\Omega^{(3)}_{ABS} = \Omega^{(3)}_{ACS} = \Omega^{(3)}_{BCS} = \pi/2$, $\Omega^{(3)}_{ABC} = 3\pi/2$, $\Omega^{(4)}_{ABCSA} = \Omega^{(4)}_{BCASB} = \Omega^{(4)}_{CABSC} = \pi$ and the basic function and the vertices read

$$\Psi_{BCC_2}(\mathbf{r}) = 2\{\cos(x+y) + \cos(x+z) + \cos(z+y) - \sin(x-y) - \sin(y-z) - \sin(z-x)\} \quad (2.12a)$$

$$\alpha_{BCC_2} = 0, \quad \beta_{BCC_2} = [\lambda_0(0) + 8\lambda_0(1) + 2\lambda_0(2) - 4\lambda(1,2)]/24. \quad (2.13a)$$

Note that the cubic vertex vanishes since for this choice of the phase shifts $\varphi_i$ the contributions of the clockwise and counterclockwise circuits for each triangle cancel each other.

The **BCC$_2$ morphology** ($I4_132$) is rather interesting and deserves some additional discussion.

i) It belongs to the class of morphologies where some of the main harmonics have a phase shift with respect to other, which, generally, leads to a decrease of the effective forth vertex. E.g., Leibler[2] already found the negative contributions into the quadric vertex under certain choice of phases for the FCC morphology:

$$\alpha_{FCC} = 0, \quad \beta_{FCC} = [\lambda_0(0) + 6\lambda_0(4/3) - 2\lambda(4/3,4/3)]/16. \quad (2.14)$$

Two more (of quite a few) representatives of this class are discussed a bit further.

ii) The shift $x = X + \pi/4$, $y = Y + \pi/4$, $z = Z + \pi/4$ of the origin of the co-ordinate system reduces the basic function (2.12a) to the form

$$\Psi_{BCC_2}(\mathbf{r}) = 4(\sin y \cos z + \sin z \cos x + \sin x \cos y). \quad (2.15)$$

The level surfaces (2.15) were considered in ref 47 as representatives of the triple periodic bicontinuous surfaces having symmetry of the $I4_132$ space group (No.214). In particular, the surface

$$\sin y \cos z + \sin z \cos x + \sin x \cos y = 0 \quad (2.15a)$$

is well known as the *Gyroid* surface (it appears rather similar to the Schoen's gyroid minimal surface[47]). So, we will also refer to the BCC$_2$ lattice as the *single gyroid*.

iii) the BCC$_2$ belongs to the class of morphologies like L, FCC, SC and SS (simple square or tetragonal cylinders[43]) we refer to as the degenerate ones because for them the cubic vertex (2.5) equals zero identically due to the symmetry reasons. For these degenerate morphologies the free energy (2.4) reads[2]

$$\Delta F_\Re = -\tau^2/(4\beta_\Re). \quad (2.4a)$$

It follows from (2.4a) that the most thermodynamically advantageous degenerate morphology is that having the least value of the quadric vertex. For references, we present the expressions for these vertices also for L, SC and SS morphologies:[2]

$$\beta_L = \lambda_0(0)/4, \quad \beta_{SC} = [\lambda_0(0) + 4\lambda_0(2)]/12, \quad \beta_{SS} = [\lambda_0(0) + 2\lambda_0(2)]/8. \quad (2.16)$$

It follows from (2.16) for $\lambda_0(0) < 2\lambda_0(2)$ and $\lambda_0(0) > 2\lambda_0(2)$ the stable morphologies are the L and SC, respectively, whereas the simple square (tetragonal planar) lattice is always metastable.

iv) The expansion (2.12a) of the corresponding basic function contains both cosines and sines, which is known[48] to imply that the BCC$_2$ morphology is non-centrosymmetric. Thus, it seems to be the simplest (and the only up to now) cubic non-centrosymmetric morphology that could be described (and for some case predicted as shown below) within the WS theory.

Now we turn to the **G family**, which is generated by the $2 \cdot 12$ main harmonics given by the 12 vectors

$$\begin{aligned}
&\mathbf{q}_{01} = (q_*/\sqrt{6})(-2,+1,+1), \quad \mathbf{q}_{02} = (q_*/\sqrt{6})(+1,-2,+1), \quad \mathbf{q}_{03} = (q_*/\sqrt{6})(+1,+1,-2),\\
&\mathbf{q}_{11} = (q_*/\sqrt{6})(-2,-1,-1), \quad \mathbf{q}_{12} = (q_*/\sqrt{6})(+1,+2,-1), \quad \mathbf{q}_{13} = (q_*/\sqrt{6})(+1,-1,+2),\\
&\mathbf{q}_{21} = (q_*/\sqrt{6})(+2,+1,-1), \quad \mathbf{q}_{22} = (q_*/\sqrt{6})(-1,-2,-1), \quad \mathbf{q}_{23} = (q_*/\sqrt{6})(-1,+1,+2),\\
&\mathbf{q}_{31} = (q_*/\sqrt{6})(+2,-1,+1), \quad \mathbf{q}_{32} = (q_*/\sqrt{6})(-1,+2,+1), \quad \mathbf{q}_{33} = (q_*/\sqrt{6})(-1,-1,-2).
\end{aligned} \quad (2.17)$$

and 12 opposite ones. The planar mapping of the vectors (2.17) is shown in Figure 2a. We use the mappings shown in Figures 1 and 2 in section 6 to derive the extinction rules for the different lattices. For comparison, in Figure 2b the planar mapping of the regular icosahedron is shown. Obviously, the set of vectors (2.17) (and the opposite ones) is obtained via a deformation of the regular icosahedron, which involves removing 6 of 30 edges of the icosahedron and the proper rotations of the left edges, the resulting polyhedron being the Wigner-Zeitz cell of the corresponding crystal lattice. It is this relationship between the G family and the icosahedron symmetry which causes the famous 10 spot SAXS pattern observed in the gyroid phase, which is one of at least three morphologies belonging to this family.

i) The morphology arising if all the phases are set to equal zero we call the **BCC$_3$**. It is just the ordinary BCC but the fact that the dominant harmonics correspond here to the 3$^{rd}$ (rather than the 1$^{st}$!) co-ordination sphere. For this morphology

$$\begin{aligned}
&\alpha_{BCC_3} = \gamma/3^{3/2},\\
&\beta_{BCC_3} = [\Lambda_1 - 8\lambda_0(1/3) - 16\lambda_0(2/3) + 4[\lambda(1/3,2/3) + \lambda(2/3,5/3)] + 2\lambda(2/3,2/3)]/48,\\
&\Lambda_1 = \lambda_0(0) + 2(\lambda_0(4/3) + 2(\lambda_0(1/3) + \lambda_0(2/3) + \lambda_0(1) + 2\lambda_0(5/3)))
\end{aligned} \quad (2.18)$$

ii) The trial function (2.3a) with the main harmonics (2.17) and the phase choice

$$\varphi_{12} = \varphi_{23} = \varphi_{31} = \varphi_{01} = \varphi_{02} = \varphi_{03} = 0, \quad \varphi_{21} = \varphi_{32} = \varphi_{13} = \varphi_{11} = \varphi_{22} = \varphi_{33} = \pi. \quad (2.19a)$$

corresponds to the bi-continuous gyroid (**G**) or *double gyroid* morphology having the symmetry $Ia\bar{3}d$. For the G phase

$$\alpha_G = \gamma/3^{3/2}, \quad \beta_G = [\Lambda_1 - 2\Lambda_2 - 4\lambda(1/3,2/3)]/48, \quad \Lambda_2 = 2\lambda(2/3,5/3) - \lambda(2/3,2/3). \quad (2.18a)$$

It is seen from (2.18), (2.18a) that due to the symmetry of the BCC$_3$ and G lattices their cubic vertices are identical and, therefore, the BCC$_3$ - G phase transition line (surface) is determined by equation

$$\beta_G - \beta_{BCC3} = 0 \quad (2.20)$$

iii) The trial function (2.3a) with the main harmonics (2.17) and the phase choice

$$\varphi_{21} = \varphi_{32} = \varphi_{13} = \pi, \ \varphi_{12} = \varphi_{23} = \varphi_{31} = \varphi_{01} = \varphi_{02} = \varphi_{03} = \varphi_{11} = \varphi_{22} = \varphi_{33} = 0 \quad (2.19b)$$

corresponds to the morphology of the symmetry $I\overline{4}3d$ we refer to as the **G$_2$**. For this morphology

$$\alpha_{G_2} = \gamma/(2 \cdot 3^{3/2}), \quad \beta_{G_2} = [\Lambda_1 + 2\Lambda_2 - 4\lambda(1/3,2/3)]/48. \quad (2.18b)$$

For reference, we give here also the vertices for the HEX morphology[2]

$$\alpha_{HEX} = 2\gamma/3^{3/2}, \quad \beta_{HEX} = [\lambda_0(0) + 4\lambda_0(1)]/12. \quad (2.21)$$

**The phase transitions lines.** The next step is to compare the free energies of the different morphologies and find at which temperatures phase transitions between the different morphologies occur. It is easy to see[2,11] from (2.4) that the free energy of an ordered phase $\Re$ becomes negative (and, thus, below the free energy of the disordered phase, which equals zero) if

$$\tilde{\tau} < \tilde{\tau}_{DIS/\Re} = (8/9)\kappa_\Re, \quad \kappa_\Re = C_\Re^2/\tilde{\beta}_\Re, \quad (2.22)$$

where the reduced temperature $\tilde{\tau} = 32\beta_0 \tau/(9\gamma^2(1))$ and the quadric vertex $\tilde{\beta}_\Re = \beta_\Re/\beta_0$ are introduced, $\beta_0$ being the minimal value of the quadric vertex for the class of the degenerate morphologies (L, BCC$_2$, FCC, SC etc.). Thus, the *first* ordered phase appearing from the disordered state is that non-degenerate phase, which has maximal value of the ratio $\kappa_\Re$. On the contrary, the *last* ordered phase appearing close to the critical point with the temperature decrease is usually (but not always!) the degenerate phase with $\tilde{\beta}_\Re = \beta_0$ we label further by the index *0*. It follows from (2.4), (2.4a) that the reduced temperature, at which a non-degenerate phase $\Re$ becomes less thermodynamically advantageous than the degenerate one, is determined as the root of the equation

$$g(\tilde{\tau}_{0/\Re}/\kappa_\Re) = \tilde{\beta}_\Re, \quad g(x) = (\sqrt{1-x}+1)(\sqrt{1-x}-(1/3))/(\sqrt{1-x}-1)^2 \quad (2.23)$$

At last, the reduced temperature of the phase transition between two non-degenerate morphologies $\Re_1$ and $\Re_2$ is the root of the equation

$$g(\tilde{\tau}_{\Re_1/\Re_2}/\kappa_{\Re_1})/\tilde{\beta}_{\Re_1} = g(\tilde{\tau}_{\Re_1/\Re_2}/\kappa_{\Re_2})/\tilde{\beta}_{\Re_2} \quad (2.24)$$

Eqs (2.22)-(2.24), which were (up to designations) presented by Leibler,[2] determine the curvatures at the critical point of the parabolic dependences of the phase transitions lines on the plane $(\tau, \gamma(1))$:

$$\tau_{\mathfrak{R}_1/\mathfrak{R}_2} = 9\tilde{\tau}_{\mathfrak{R}_1/\mathfrak{R}_2}\,\gamma^2(1)/(32\beta_0). \tag{2.25}$$

Thus, solving eqs (2.22)-(2.22) successively provides us with the information concerning the existence and the order of appearance of the stable morphologies at the very critical point. Therewith, the explicit form of the function $\lambda(h_1, h_2)$, which appears in the expression (2.6) for the fourth vertex $\beta_{\mathfrak{R}}$, should be taken into account to determine the final shape of the phase diagrams. Following Leibler[2] we refer to the $h$-dependence of the vertex $\lambda$ as the angle one since the values of the parameters $h_i$ depend on the angles between the vectors $\mathbf{q}_i$.

For diblock copolymers the angle dependence is rather weak[2], which enabled Fredrickson and Helfand[27] to propose the following commonly accepted approximation:

$$\lambda(h_1, h_2) \approx \lambda(0,0) = \lambda_0(0). \tag{2.26}$$

However, it cannot be expected to be (and, as shown further, is not) true for any systems. So, it is instructive first to analyze the effects of this angle dependence on the phase diagram of the weakly crystallizing systems in the simplest assumption[44]

$$\lambda(h_1, h_2, h_3) = \sum_{i=1}^{3} f(h_i) = \lambda_0\left(1 - \frac{3\delta}{32}\left(4^2 - \sum_{i=1}^{3} h_i^2\right)\right). \tag{2.27}$$

Such a choice is due to the fact that i) the expression (2.27) keeps only the first non-constant term in the expansion of $\Gamma_4$ in powers of $h_i$, and ii) the positive (negative) sign of $\delta$ corresponds to a disadvantage (advantage) of the lamellar structure as compared to all other ones (in the original ref 44 we used the opposite choice for the sign of the parameter $\delta$).

The phase diagram describing the phase transition lines starting at the very critical point on the plane ($\delta$, the reduced temperature $\tilde{\tau} = 32\tau\lambda_0/(9\gamma^2)$) is presented in Figure 3 (in the original ref 44 it was build within the interval $0 \leq |\delta| \leq 1$ only). It is seen that the conventional phase transition sequence DIS-BCC-HEX-LAM occurs only for $\delta<\delta_0=0.362$. On the contrary, for $\delta>\delta_0$ the following non-conventional sequences occur:

i) the sequence DIS-BCC-HEX-G-LAM for $\delta_{12} > \delta > \delta_0$, $\delta_{12} = 4/9$;

ii) the sequence DIS-BCC-HEX-G-BCC$_2$ for $\delta_{23} > \delta > \delta_{12}$, $\delta_{23} = 2/3$;

iii) the sequence DIS-BCC-HEX-G-FCC for $\delta_1 > \delta > \delta_{23}$, $\delta_1 = 0.822$;

iv) the sequence DIS-BCC-G-FCC for $\delta_{34} > \delta > \delta_1$, $\delta_{34} = 5/6$;

v) the sequence DIS-BCC-G-SC for $\delta_{45} > \delta > \delta_{34}$, $\delta_{45} = 0.891$;

vi) the sequence DIS-BCC-SC for $\delta_2 > \delta > \delta_{45}$, $\delta_2 = 4/3$;

vii) the sequence DIS-BCC for $\delta_{lim} > \delta > \delta_2$, $\delta_{lim} = 1.538$;

At last, $\beta_{BCC}$ becomes negative for $\delta > \delta_{lim}$, in which case one should take into account the next (5$^{th}$, 6$^{th}$ etc.) terms of the Landau expansion. Thus, the original weak segregation approximation is not valid anymore for $\delta > \delta_{lim}$.

The crucial question for the preceding analysis to be relevant is whether there are any real substances where the angle dependence is strong enough to cause some non-conventional morphology to exist. For long time we tried to find such a real substance among molten binary AB copolymers of various architectures basing on a Leibler microscopic theory of weak segregation.[2,19] However, these attempts went unrewarded until we turned to the ternary ABC copolymers.[45,46] (See also refs 51 where the non-conventional morphologies were found in some properly designed ABC-like binary block copolymers basing on the analysis given in ref 46 and in more detail below.) So, in the next section we give the complete WS description of the ternary ABC copolymers.

**3. Weak segregation description of the ternary ABC copolymers: weakly and strongly fluctuating order parameters and scalar parameterization of the free energy near the ODT.**

The order parameter for these systems becomes a 3-component vector and the Landau free energy takes the form[43,52]

$$F = T\int \frac{d\mathbf{r}}{v}\left(\chi_{AB}\phi_A(\mathbf{r})\phi_B(\mathbf{r}) + \chi_{BC}\phi_B(\mathbf{r})\phi_C(\mathbf{r}) + \chi_{AC}\phi_A(\mathbf{r})\phi_C(\mathbf{r})\right) + F_{str}(\{\phi_i(\mathbf{r})\}). \quad (3.1)$$

Here $T$ is the temperature measured in the energetic units ($k_B = 1$), the first term is the interaction contribution evaluated for the Flory-Huggins lattice model and the second is so-called structural free energy $F_{str}$ corresponding to the entropy of the inhomogeneous ideal system of the copolymer macromolecules under consideration with certain specified spatial profiles of the volume fractions $\phi_i(\mathbf{r})$ of the repeated units of the *i*-th sort. (For simplicity, we assume that the excluded volume for all sorts of the repeating units (monomers) is the same and equal $v$.) Within the WS theory $F_{str}$ is evaluated by its expansion in powers of the local deviations $\Phi_i(\mathbf{r}) = \phi_i(\mathbf{r}) - \overline{\phi}_i$ ($\phi_1 = \phi_A$, $\phi_2 = \phi_B$, $\phi_3 = \phi_C$) of the volume fractions from their value averaged over the whole volume of the system ($\overline{\phi}_i = f_i$ for the incompressible melt we restrict ourselves to in this paper):

$$F_{str}(\{\phi_i(\mathbf{r})\}) = F_{str}(\{f_i\}) + \Delta F_2 + \Delta F_3 + \Delta F_4 + \ldots. \tag{3.2}$$

where the contributions $\Delta F_n$ are defined as follows:

$$\frac{\Delta F_2}{T} = \frac{1}{2}\int \frac{(\mathbf{g}^{-1}(q))_{ij} \Phi_i(\mathbf{q})\Phi_j(-\mathbf{q})d\mathbf{q}}{(2\pi)^3}, \tag{3.3}$$

$$\frac{\Delta F_n}{T} = \frac{1}{n!}\int \delta\left(\sum_{i=1}^n \mathbf{q}_i\right) \Gamma^{(n)}_{\alpha_1,\ldots,\alpha_n}(\mathbf{q}_1,\ldots,\mathbf{q}_n) \prod_{i=1}^n \frac{\Phi_{\alpha_i}(\mathbf{q}_i)d\mathbf{q}_i}{(2\pi)^3}, \quad n = 3,4 \tag{3.4}$$

The structural matrix $\mathbf{g}(q)$ in (3.3) was introduced in ref 53 (see also ref 54). This matrix as well as the 3-dimensional tensors $\Gamma^{(3)}$ and $\Gamma^{(4)}$ of the rank 3 and 4, respectively, in (3.4) are related to the architecture of the system as described in refs 52, 55.

The terms of the higher order in power of $\Phi$ in the infinite series (3.2) are skipped since the weak segregation approximation we restrict ourselves here is the Landau-type theory where the terms up to the forth order in $\Phi$ are only relevant. In the definition (3.3) and thereafter we imply summation over all the repeating indices running the values *1,...,3*.

It is convenient to use explicitly the incompressibility condition

$$\Phi_A(\mathbf{r}) + \Phi_B(\mathbf{r}) + \Phi_C(\mathbf{r}) = 0, \tag{3.5}$$

Substituting eqs (3.2)-(3.5) into the total free energy (3.1) one can rewrite it as a functional of two only (rather than three) order parameters $\Phi_1(\mathbf{r}) = \Phi_A(\mathbf{r})$ and $\Phi_2(\mathbf{r}) = \Phi_C(\mathbf{r})$:

$$F(\{\phi_i(\mathbf{r})\}) = F(\{f_i\}) + \Delta F_2 + \Delta F_3 + \Delta F_4 + \ldots \tag{3.6}$$

Now the contributions $\Delta F_n$ are defined as follows:

$$\frac{\Delta F_2}{T} = \frac{1}{2}\int \frac{d^3q}{(2\pi)^3} \sum_{i,j=1}^{2}(\mathbf{G}^{-1}(q))_{ij}\Phi_i\Phi_j, \tag{3.7}$$

$$\frac{\Delta F_n}{T} = \frac{1}{n!}\int \delta\left(\sum_{i=1}^{n} \mathbf{q}_i\right) \tilde{\Gamma}^{(n)}_{\alpha_1,...,\alpha_n}(\mathbf{q_1},..,\mathbf{q}_n) \prod_{i=1}^{n} \frac{\Phi_{\alpha_i}(\mathbf{q}_i) d\mathbf{q}_i}{(2\pi)^3}, \quad n=3,4 \qquad (3.8)$$

Here[57]

$$\mathbf{G}^{-1} = \begin{pmatrix} (\overline{\mathbf{g}}^{-1})_{11} - 2\chi_{AB} & (\overline{\mathbf{g}}^{-1})_{12} + \chi_{AC} - \chi_{AB} - \chi_{BC} \\ (\overline{\mathbf{g}}^{-1})_{12} + \chi_{AC} - \chi_{AB} - \chi_{BC} & (\overline{\mathbf{g}}^{-1})_{22} - 2\chi_{BC} \end{pmatrix}, \qquad (3.9)$$

$$\overline{\mathbf{g}}^{-1} = \begin{pmatrix} (\mathbf{g}^{-1})_{AA} - 2(\mathbf{g}^{-1})_{AB} + (\mathbf{g}^{-1})_{BB} & (\mathbf{g}^{-1})_{AC} + (\mathbf{g}^{-1})_{BB} - (\mathbf{g}^{-1})_{AB} - (\mathbf{g}^{-1})_{BC} \\ (\mathbf{g}^{-1})_{AC} + (\mathbf{g}^{-1})_{BB} - (\mathbf{g}^{-1})_{AB} - (\mathbf{g}^{-1})_{BC} & (\mathbf{g}^{-1})_{CC} - 2(\mathbf{g}^{-1})_{CB} + (\mathbf{g}^{-1})_{BB} \end{pmatrix}, \qquad (3.10)$$

where the components of the matrix $\mathbf{g}(q)$ are calculated for ABA copolymers in refs 57, 58 and for ABC copolymers in refs 59, 60, and

$$\begin{cases} \tilde{\Gamma}^{(3)}_{\alpha\beta\gamma}(\mathbf{q_1},..,\mathbf{q}_n) = \Gamma^{(3)}_{ijk}(\mathbf{q_1},..,\mathbf{q}_n) c_{i\alpha} c_{j\beta} c_{k\gamma}, \\ \tilde{\Gamma}^{(4)}_{\alpha\beta\gamma\delta}(\mathbf{q_1},..,\mathbf{q}_n) = \Gamma^{(4)}_{ijkl}(\mathbf{q_1},..,\mathbf{q}_n) c_{i\alpha} c_{j\beta} c_{k\gamma} c_{l\delta}, \end{cases} \mathbf{c} = \begin{pmatrix} 1 & -1 & 0 \\ 0 & -1 & 1 \end{pmatrix}, \begin{cases} i,j,k,l=1,2,3; \\ \alpha,\beta,\gamma,\delta=1,2 \end{cases} \qquad (3.11)$$

Thus, the main new feature of the incompressible *n*-component systems is that they are described by $(n-1)^3$ third vertices rather than one. In fact, due to the symmetry reasons there are four independent third vertices ($g_{111}$, $g_{222}$, $g_{112}$, $g_{221}$) for the *3*-component systems even in the most general case. Anyway, having in mind that the WS theory holds, by definition, in a vicinity of the critical point (i.e. that where all the cubic terms vanish), it could appear[43] that it is hardly possible at all to apply the WS theory to many-component systems. Nevertheless, as shown by the author and Dobrynin[55] via an extension of the analysis given in refs 52, 61, it is possible to reduce consideration of *n*-component systems to the form presented in the previous section for those with a scalar order parameter.

We present such a reduction for an arbitrary number *n* of independent order parameters and start it with a reminder that the uniform (disordered) state of the systems is defined as that where all the components of the vector order parameter vanish:

$$\Phi_i(\mathbf{r}) = 0, \quad i = 1, n \qquad (3.12)$$

Within the region of the disordered state stability (at least, metastability), the quadratic form $\Gamma_{ij}(\mathbf{q})\Phi_i(\mathbf{q})\Phi_j(-\mathbf{q})$ is positive definite. Then all the eigenvalues $\lambda_s(q), s=1,..,n$ of the matrix $\boldsymbol{\Gamma} = \left\| \Gamma_{ij}(\mathbf{q}) \right\|$ are positive for any $q = |\mathbf{q}|$, $0 \leq q \leq \infty$. So, the matrix $\Gamma$ is non-degenerate, in which case the following matrix identities hold:

$$\left\| \Gamma_{ij}(\mathbf{q}) \right\| = \Lambda(q) E_i(q) E_j(q) + \sum_{s=2}^{n} \lambda_s(q) e_i^{(s)}(q) e_i^{(s)}(q), \tag{3.13a}$$

$$\mathbf{I} = \left\| \delta_{ij} \right\| = E_i(q) E_j(q) + \sum_{s=2}^{n} e_i^{(s)}(q) e_j^{(s)}(q), \tag{3.13b}$$

where $\mathbf{I}$ is the identity matrix, $\Lambda(q)$ and $\lambda_i(q)$ ($i=2,...n$) are the minimal and all other eigenvalues of the matrix $\Gamma$, respectively, and $E_i(q)$ and $e_i^{(s)}(q)$ are the corresponding eigenvectors, all of which are some functions of the wave number $q$ (in this paper we assume that there is only one minimal eigenvector for every value of $q$). We refer to the projections of the vector order parameter $\{\Phi_i(\mathbf{q})\}$ onto the corresponding eigenvectors

$$\Psi(\mathbf{q}) = E_i(q) \Phi_i(\mathbf{q}), \quad \varphi_s(\mathbf{q}) = e_i^{(s)}(q) \Phi_i(\mathbf{q}), \quad s = 2,..,n. \tag{3.14}$$

as the strongly and weakly fluctuating fields, respectively. The free energy (3.6) can be rewritten in terms of these fields as follows:

$$\Delta F(\{\Psi\}, \{\varphi_s\}) = \Delta F_{\text{strong}}(\{\Psi\}) + \Delta F_{\text{weak}}(\{\varphi_s\}) + \Delta F_{\text{coupling}}(\{\Psi, \varphi_s\}). \tag{3.15}$$

Here

$$\frac{\Delta F_{\text{strong}}}{T} = \frac{1}{2} \int \Lambda(q) |\Psi(\mathbf{q})|^2 \frac{d\mathbf{q}}{(2\pi)^3} + \frac{1}{3!} \int \delta\left(\sum_{i=1}^{3} \mathbf{q}_i\right) \Gamma_3(\mathbf{q}_1,..,\mathbf{q}_n) \prod_{i=1}^{3} \frac{\Psi(\mathbf{q}_i) d\mathbf{q}_i}{(2\pi)^3}$$
$$+ \frac{1}{4!} \int \delta\left(\sum_{i=1}^{4} \mathbf{q}_i\right) \beta_{\text{strong}}(\mathbf{q}_1,..,\mathbf{q}_4) \prod_{i=1}^{4} \frac{\Psi(\mathbf{q}_i) d\mathbf{q}_i}{(2\pi)^3}, \tag{3.16}$$

is the contribution to the total free energy of a system characterized by a non-uniform profile $\{\Phi_i(\mathbf{q})\}$ of the vector order parameter if all the weakly fluctuating fields are zero,

$$\frac{\Delta F_{\text{weak}}}{T} = \frac{1}{2} \int \sum_{s=2}^{n} \lambda_s(q) |\varphi_s(\mathbf{q})|^2 \frac{d\mathbf{q}}{(2\pi)^3} + ... \tag{3.17}$$

is that if the strongly fluctuating field is arbitrarily set to equal zero, and the term

$$\frac{\Delta F_{\text{coupling}}}{T} = \frac{1}{2!} \sum_{s=2}^{n} \int \delta\left(\sum_{i=1}^{3} \mathbf{q}_i\right) \tilde{\gamma}_i^{(1)}(\mathbf{q}_1,..,\mathbf{q}_3) \, \varphi_s(\mathbf{q}_1) \Psi(\mathbf{q}_2) \Psi(\mathbf{q}_3) \prod_{i=1}^{3} \frac{d\mathbf{q}_i}{(2\pi)^3} + ... \tag{3.18}$$

is the field coupling contribution. The vertices appearing explicitly in expressions (3.16)-(3.18) read:

$$\Gamma_3(\mathbf{q_1},..,\mathbf{q_3}) = \Gamma^{(3)}_{\alpha_1,\alpha_2,\alpha_3}(\mathbf{q_1},..,\mathbf{q_3}) E_{\alpha_1}(\mathbf{q_1}) E_{\alpha_2}(\mathbf{q_2}) E_{\alpha_3}(\mathbf{q_3}), \tag{3.19a}$$

$$\beta_{strong}(\mathbf{q_1},..,\mathbf{q_4}) = \Gamma^{(4)}_{\alpha_1,\alpha_2,\alpha_3,\alpha_4}(\mathbf{q_1},..,\mathbf{q_4}) E_{\alpha_1}(\mathbf{q_1}) E_{\alpha_2}(\mathbf{q_2}) E_{\alpha_3}(\mathbf{q_3}) E_{\alpha_4}(\mathbf{q_4}). \tag{3.19b}$$

$$\tilde{\gamma}^{(1)}_i(\mathbf{q_1},..,\mathbf{q_3}) = \Gamma_{\alpha_1,\alpha_2,\alpha_3}(\mathbf{q_1},..,\mathbf{q_3}) e^i_{\alpha_1}(\mathbf{q_1}) E_{\alpha_2}(\mathbf{q_2}) E_{\alpha_3}(\mathbf{q_3}). \tag{3.19c}$$

The reason to do this (at the first look threateningly cumbersome) change of the variables is that it enables us to estimate easily the orders of magnitude of all the contributions and keep only those, which do not exceed the accuracy of the WS theory. Therewith, all the terms we skipped when writing the contributions of the inter-field coupling and weakly fluctuating fields turn out to exceed this accuracy, which enables us to neglect them.

Indeed, similar to consideration in the preceding section the system morphology is described by the non-zero vector order parameter $\{\Phi_i(\mathbf{r})\}$ profile providing the minimum of the free energy (3.6). It is convenient to find this minimum in two steps. First, we fix a finite profile of the strongly fluctuating field $\Psi(\mathbf{q})$ and minimize the total free energy (3.6) with respect to the fields $\{\varphi_s(\mathbf{q})\}$. The result of such a minimization is a functional of the scalar field $\Psi(\mathbf{q})$ only:

$$\Delta F_{eff}(\{\Psi\}) = \Delta F_{strong}(\{\Psi\}) + \Delta \tilde{F}(\{\Psi(\mathbf{q})\}), \tag{3.20}$$

where

$$\Delta \tilde{F}(\{\Psi(\mathbf{q})\}) = \min_{\{\varphi_s\}} \{\Delta F_{weak}(\{\varphi_s\}) + \Delta F_{coupling}(\{\Psi, \varphi_s\})\} \tag{3.21}$$

The next step is to minimize the effective free energy $\Delta F_{eff}(\{\Psi\})$ with respect to the scalar field $\Psi(\mathbf{q})$. If the coupling contribution is described by the only term explicitly written in (3.18), the functional $\Delta F_{eff}(\{\Psi\})$ is calculated straightforwardly and the corresponding minimum condition and the result of the minimization read

$$\varphi_s(\mathbf{q}) = -(2\lambda_s)^{-1} \int \tilde{\gamma}^{(1)}_s(-\mathbf{q},\mathbf{p},\mathbf{q}-\mathbf{p}) \Psi(\mathbf{p})\Psi(\mathbf{q}-\mathbf{p}) d\mathbf{p}/(2\pi)^3, \tag{3.21}$$

$$\frac{\Delta \tilde{F}(\{\Psi(\mathbf{q})\})}{T} = -\frac{1}{4!} \int \delta\left(\sum_{i=1}^n \mathbf{q}_i\right) \Delta\beta(\mathbf{q_1},..,\mathbf{q_4}) \prod_{i=1}^4 \frac{\Psi(\mathbf{q}_i) d\mathbf{q}_i}{(2\pi)^3} \tag{3.22}$$

where

$$\Delta\beta(\mathbf{q_1},..,\mathbf{q_4}) = B(\mathbf{q_1},\mathbf{q_2};\mathbf{q_3},\mathbf{q_4}) + B(\mathbf{q_1},\mathbf{q_3};\mathbf{q_2},\mathbf{q_4}) + B(\mathbf{q_1},\mathbf{q_4};\mathbf{q_2},\mathbf{q_3}), \tag{3.23a}$$

$$B(\mathbf{q_1},\mathbf{q_2};\mathbf{q_3},\mathbf{q_4}) = \sum_{s=2}^{N} \tilde{\gamma}_s^{(1)}(\mathbf{q_1},\mathbf{q_2},\mathbf{p}) \lambda_s^{-1}(|\mathbf{p}|) \tilde{\gamma}_s^{(1)}(\mathbf{q_3},\mathbf{q_4},-\mathbf{p}), \quad \mathbf{p} = -\mathbf{q_1} - \mathbf{q_2} = \mathbf{q_3} + \mathbf{q_4}. \tag{3.23b}$$

As is seen from eqs (3.21), (3.22), a small strongly fluctuating field $\Psi$ induces weakly fluctuating fields $\varphi_s$, which with the coupling $\varphi_s \Psi^2$ are of the order of magnitude $O(\Psi^2)$. As the result, an extra (as compared to $\Delta F_{\text{strong}}(\{\Psi\})$) contribution of the order $O(\Psi^4)$ occurs in the free energy of the system with a fixed value of $\Psi$. All other terms skipped in (3.17), (3.18) are of the order $\Psi\varphi^2 \sim \Psi^5$, $\varphi^3 \sim \Psi^6$, $\Psi^2\varphi^2 \sim \Psi^6$ and higher orders of magnitude with respect to the small quantity $\Psi$. So, we would have to keep them in the free energy only if we kept simultaneously the terms of the same order $\Psi^5, \Psi^6$ and higher that are present in the expansion of the strongly fluctuating field contribution $\Delta F_{\text{strong}}(\{\Psi\})$. But since we restrict ourselves to the weak segregation approximation (Landau-type expansion up to the forth order of $\Psi$) we are to neglect any terms of the order $o(\Psi^4)$ whatever is their origin.

It follows from eqs. (3.6), (3.10), (3.12) that the final effective free energy reads

$$\frac{\Delta F_{\text{eff}}(\{\Psi(\mathbf{q})\})}{T} = \frac{1}{2} \int \Lambda(q) |\Psi(\mathbf{q})|^2 \frac{d\mathbf{q}}{(2\pi)^3} + \frac{1}{3!} \int \delta\left(\sum_{i=1}^{n} \mathbf{q}_i\right) \Gamma_3(\mathbf{q_1},..,\mathbf{q}_n) \prod_{i=1}^{3} \frac{\Psi(\mathbf{q}_i) d\mathbf{q}_i}{(2\pi)^3}$$
$$+ \frac{1}{4!} \int \delta\left(\sum_{i=1}^{n} \mathbf{q}_i\right) \Gamma_4^{\text{eff}}(\mathbf{q_1},..,\mathbf{q_4}) \prod_{i=1}^{4} \frac{\Psi(\mathbf{q}_i) d\mathbf{q}_i}{(2\pi)^3} \tag{3.24}$$

It looks just like the free energy (2.1) with the scalar field $\Psi$ where the cubic vertex is defined by eq (3.19a) and forth vertex reads

$$\Gamma_4^{\text{eff}}(\mathbf{q_1},..,\mathbf{q_4}) = \beta_{\text{strong}}(\mathbf{q_1},..,\mathbf{q_4}) - \Delta\beta(\mathbf{q_1},..,\mathbf{q_4}). \tag{3.25}$$

The spinodal condition determining stability of an $n$-component system uniform state is[3]

$$\Lambda(q_*) = \min \Lambda(q) = 0, \tag{3.26}$$

where $q = q_*$ is the location of the minimum of the function $\Lambda(q)$. The situations $q_* = 0$ and $q_* > 0$ correspond to the instabilities as to macro- and microphase separation, respectively.

Comparing the designations used in (3.24) with those of the preceding section, we see that the location of the critical point is given by vanishing of the only scalar combination of the components of the tensor of the third vertices:

$$\gamma(1) = 0. \tag{3.27}$$

Here the function $\gamma(h)$ is related to the cubic vertex (3.19a) via eq (2.8). $(n-1)$ more scalar combinations $\gamma_i^{(1)}$ of the components of the tensor of the third vertices, which appear in the coupling term (3.18) and are defined in (3.19c), contribute into the renormalized forth vertex as consistent with (3.23). All other ones are irrelevant since they appear only in the contributions exceeding the WS theory accuracy, which is of order of $o(\Psi^4)$.

Thus, we did reduce the problem of minimization of the Landau free energy (3.1) with $n$-vector order parameter to that of the effective free energy (3.24) with a scalar one. The next step would be to solve the problem explicitly in the WS approximation as it was shown in the section 2, i.e. by substituting into (3.24) the trial function (2.3a):

$$\phi(\mathbf{r}) = A\Psi(\mathbf{r}), \quad \Psi(\mathbf{r}) = \frac{1}{2}\sum_{|\mathbf{q}_i|=q_*}\{\exp i(\mathbf{q}_i\mathbf{r}+\varphi_i)+c.c\} \tag{3.28}$$

However, during the last decade some authors[50,62-64] queried reliability of the approximation (3.28). In particular, Hamley and Podnek[64] suggested that the gyroid morphology existence is due to anomalously large (negative) contribution to the total free energy of the second harmonics with $h = \mathbf{q}^2/q_*^2 = 4/3$ characteristic for the G morphology.

Of course, this suggestion by itself is not sufficient to explain the G phase stability in diblock copolymers since the second harmonics with $h = 4/3$ are characteristic also for the $BCC_3$ and $G_2$ morphologies, which all belong to the G family. Moreover, it has no relation to the problem of the G stability at the very critical point, which depends only on the strength of the angle dependence of the forth vertex as shown above. However, if the angle dependence is not strong enough to provide the G stability at the critical point, we are to deal with two closely related problems: *i)* which phases are stable at the triple points (if any) existing near the critical point; and *ii)* which factors determine location of the triple points. In general, contribution of *many* harmonics (rather than that of the only second ones) determines location of the triple point in question, which is shown via direct calculation by Matsen and Schick.[24,25] But it is natural to expect that only certain finite number of the higher harmonics is relevant if the triple points are close enough to the critical one.

The contributions of such selected higher harmonics could be included into the general scheme of the WS theory via a straightforward generalization of the procedure we used in the preceding section to separate contributions of the weakly and strongly fluctuating order parameters. Namely, we chose the trial function as follows:

$$\phi(\mathbf{r}) = \Psi(\mathbf{r}) + \sum_{h \neq 1} \psi_h(\mathbf{r}), \tag{3.29}$$

where the sum of the main harmonics

$$\Psi(\mathbf{r}) = A \sum_{\mathbf{q}_i \in \mathfrak{R}^{-1},\, \mathbf{q}_i^2 = q_*^2} \exp i(\mathbf{q}_i \mathbf{r} + \varphi_i) \tag{3.30a}$$

and that of the higher ones

$$\psi_h(\mathbf{r}) = \sum_{\mathbf{q}_i \in \mathfrak{R}^{-1},\, \mathbf{q}_i^2 = h q_*^2} a_{\mathbf{q}_i} \exp i\big((\mathbf{q}_i \mathbf{r}) + \varphi(\mathbf{q}_i)\big) \tag{3.30b}$$

play roles of the strongly and weakly fluctuating fields, respectively.

Now, the expression for the effective free energy (3.14) reads similarly to representation (3.5):

$$\Delta F = \Delta F_{\text{main}} + \Delta F_{\text{high}} + \Delta F_{\text{coupling}}. \tag{3.31}$$

Here

$$\Delta F_{\text{main}} = \Delta F_{\text{eff}}(\{\Psi(\mathbf{r})\}), \tag{3.32}$$

where $\Delta F_{\text{eff}}(\{\Psi\})$ is defined by eq (3.14), is the contribution of the dominant harmonics (4.2a),

$$\Delta F_{\text{high}} = \frac{VT}{2} \sum_{\mathbf{q}_i \in \mathfrak{R}^{-1},\, \mathbf{q}_i^2 = h q_*^2} \left(C(h-1)^2 + \tau\right) |a_{\mathbf{q}_i}|^2, \quad C = (1/2) \partial^2 \Lambda(q^2)/\partial(q^2)^2 \Big|_{q^2 = q_*^2}$$

$$\tag{3.33}$$

is the contribution of the higher harmonics and

$$\Delta F_{\text{coupling}} = \Delta F_{\text{coupling}}^{(2)} + \Delta F_{\text{coupling}}^{(3)} + \ldots$$

$$\tag{3.34}$$

is the contribution due to coupling between the dominant and higher harmonics, generated by the cubic and quadric terms of the original Hamiltonian:

$$\Delta F^{(2)}_{coupling} = VT \frac{A_0^2}{2} \sum_{\mathbf{q}_i \in \mathfrak{R}^{-1}, \mathbf{q}_i^2 = q_*^2} \gamma(\mathbf{q}_1, \mathbf{q}_2, -\mathbf{q}_1 - \mathbf{q}_2) a_{\mathbf{q}_1 + \mathbf{q}_2} \exp(i(\phi_1 + \phi_2)), \quad (3.35a)$$

$$\Delta F^{(3)}_{coupling} = VT \frac{A_0^3}{6} \sum_{\substack{\mathbf{q}_i \in \mathfrak{R}^{-1} \\ \mathbf{q}_i^2 = q_*^2}} \Gamma_4(\mathbf{q}_1, \mathbf{q}_2, \mathbf{q}_3, -\mathbf{q}_1 - \mathbf{q}_2 - \mathbf{q}_3) a_{\mathbf{q}_1 + \mathbf{q}_2 + \mathbf{q}_3} \exp(i(\phi_1 + \phi_2 + \phi_3)) \quad (3.35b)$$

Again, it is only $\Delta F^{(2)}_{coupling}$ term, which is relevant in the final result. Indeed, minimization of the free energy (3.31) with respect to the complex amplitudes $a_\mathbf{q}$, $\mathbf{q}^2/q_*^2 = h \neq 1$, gives

$$a_\mathbf{q} = -\frac{A_0^2}{2} \frac{D(\mathbf{q})\gamma(h)}{C(h-1)^2 + \tau}, \quad (3.36)$$

$$D(\mathbf{q}) = \sum_\mathbf{q} \exp(i(\phi_i + \phi_j)) \quad (3.37)$$

where the symbol $\Sigma_\mathbf{q}$ implies summation over all pairs of the main harmonics satisfying the condition

$$\mathbf{q}_i + \mathbf{q}_j + \mathbf{q} = 0. \quad (3.38)$$

According to the formulas (3.36)-(3.38) the second harmonics induced by coupling (3.35a), which are to be taken into account within the Landau Hamiltonian accuracy, belong to all the coordination spheres of the corresponding conjugated lattice, radius of which does not exceed the doubled radius of the dominant coordination sphere. The number of the coordination spheres of different radius satisfying this condition depends on the lattice symmetry and varies from *1* (for the LAM) to 8 (for the G) and 10 (for the BCC$_3$). We will refer to all of the corresponding harmonics as the 2$^{nd}$ shell harmonics.

For some of these harmonics the factor (3.37) could vanish identically for the lattices having non-zero phase shifts like BCC$_2$, G and G$_2$, which gives a natural derivation of the extinction rules[48] within the WS theory. We return to this important issue in more detail in section 6 where we discuss the observable small-angle scattering patterns for different non-conventional lattices.

At last, substituting (3.35a), (3.36), (3.37) into eqs (3.31)-(3.34) we see that the final expression for the free energy including the higher harmonics contribution up to order of $O(A_0^4)$ reads

$$\Delta F_{main} = VT(\tau A_0^2 + \alpha_\Re A_0^3 + \beta_\Re A_0^4), \tag{3.39}$$

where $\alpha_\Re, \beta_\Re$ are related to the cubic vertex (3.9a) and the fully renormalized forth vertex

$$\Gamma_4(\mathbf{q_1},..,\mathbf{q_4}) = \Gamma_4^{eff}(\mathbf{q_1},..,\mathbf{q_4}) - (\overline{B}(\mathbf{q_1},\mathbf{q_2};\mathbf{q_3},\mathbf{q_4}) + \overline{B}(\mathbf{q_1},\mathbf{q_3};\mathbf{q_2},\mathbf{q_4}) + \overline{B}(\mathbf{q_1},\mathbf{q_4};\mathbf{q_2},\mathbf{q_3})), \tag{3.40}$$

$$\overline{B}(\mathbf{q_1},\mathbf{q_2};\mathbf{q_3},\mathbf{q_4}) = \sum_{s=2}^{N} \gamma^2(h) / \left( C((p^2/q_*^2) - 1)^2 + \tau \right) \quad \mathbf{p} = -\mathbf{q_1} - \mathbf{q_2} = \mathbf{q_3} + \mathbf{q_4}, \tag{3.41}$$

via eqs (2.6)-(2.11).

### 4. Spinodals and critical lines in molten ternary ABC triblock and mictoarm copolymers.

Now we apply the procedure presented in the preceding section to analyze the order-disorder and order-order transition lines close to the critical points of the molten monodisperse incompressible ternary $A_nB_mC_l$ triblock and tristar (mictoarm) copolymers. The system is specified by the values of the interaction parameters $\chi_{AB}$, $\chi_{BC}$, and $\chi_{AC}$ and compositions $f_A = n/N$, $f_B = m/N$, $f_C = l/N$ ($N=n+m+l$ is the total degree of polymerization).

To begin with, let us consider the symmetric case $\chi_{AB} = \chi_{BC} = \chi$, $f_A = f_C = f$ and assume for simplicity that the Kuhn lengths of all three blocks are the same. Then $(\tilde{\mathbf{g}}^{-1})_{11}(q) = (\tilde{\mathbf{g}}^{-1})_{22}(q)$ and it follows from eqs (3.7), (3.9) and (3.10) that the quadratic term (3.7) reads

$$\Delta F_2 = \frac{1}{4N} \int \frac{d^3q}{(2\pi)^3} \left( \lambda_+ (\Phi_+(\mathbf{r}))^2 + \lambda_- (\Phi_-(\mathbf{r}))^2 \right),$$
$$\lambda_+ = (\tilde{\mathbf{g}}^{-1})_{11}(q) + (\tilde{\mathbf{g}}^{-1})_{12}(q) + \tilde{\chi}_{AC} - 4\tilde{\chi}, \quad \Phi_+(\mathbf{r}) = \Phi_A(\mathbf{r}) + \Phi_C(\mathbf{r}), \tag{4.1}$$
$$\lambda_- = (\tilde{\mathbf{g}}^{-1})_{11}(q) - (\tilde{\mathbf{g}}^{-1})_{12}(q) - \tilde{\chi}_{AC}, \quad \Phi_-(\mathbf{r}) = \Phi_A(\mathbf{r}) - \Phi_C(\mathbf{r})$$

Here we introduced the reduced $\chi$-parameters $\tilde{\chi}_{AC} = \chi_{AC}N$, $\tilde{\chi} = \chi N$ and structural matrix $\tilde{\mathbf{g}} = \overline{\mathbf{g}}/N$, components of which depend on the reduced squared wave number $Q = q^2 a^2 N/6$ and the compositions $f_i$ ($i=A,B,C$) only. (It is possible since all the vertexes $\Gamma^{(k)}$ are proportional to $1/N$.)

It follows from (4.1) that both the character of the weakly segregated morphology occurring in such a symmetric system and the very possibility to describe it within the WS theory depends crucially on the values of the interaction parameters. Namely, the plane $(\tilde{\chi}_{AC}, \tilde{\chi})$ is divided by the lines $\lambda_-(\tilde{\chi}_{AC}, \tilde{\chi}) = 0$ and $\lambda_+(\tilde{\chi}_{AC}, \tilde{\chi}) = 0$ into four regions:

*i)* the stability region ($\lambda_- > 0, \lambda_+ > 0$), where the fluctuations of both order parameters $\Phi_+(\mathbf{r}), \Phi_-(\mathbf{r})$ are finite and the uniform state is stable (or at least metastable) with respect to these fluctuations;

*ii)* the AC-modulation region ($\lambda_- < 0, \lambda_+ > 0$), where the uniform (disordered) state is unstable with respect to formation of certain profile $\Phi_-(\mathbf{r}) \neq 0$, the order parameter $\Phi_+(\mathbf{r})$ being weakly fluctuating;

*iii)* the B-modulation region ($\lambda_+ < 0, \lambda_- > 0$), where the uniform state is unstable with respect to formation of certain profile $\Phi_+(\mathbf{r}) = -\Phi_B(\mathbf{r}) \neq 0$, the order parameter $\Phi_-(\mathbf{r})$ being weakly fluctuating;

*iv)* the region ($\lambda_- < 0, \lambda_+ < 0$), where the uniform state is unstable with respect to fluctuations of both order parameters $\Phi_+(\mathbf{r})$ and $\Phi_-(\mathbf{r})$.

It follows from (4.1) that the lines $\lambda_+ = 0$ and $\lambda_- = 0$ in the plane $(\tilde{\chi}_{AC}, \tilde{\chi})$ are the straight lines

$$\chi_{AC} - 4\chi = -\min b(q) = -b(q_+), \quad b(q) = \min\left((\tilde{\mathbf{g}}^{-1})_{11}(q) + (\tilde{\mathbf{g}}^{-1})_{12}(q)\right), \tag{4.2a}$$

$$\chi_{AC} = \min a(q) = a(q_-), \quad a(q) = (\tilde{\mathbf{g}}^{-1})_{11}(q) - (\tilde{\mathbf{g}}^{-1})_{12}(q), \tag{4.2b}$$

where the critical wave numbers $q_+$ and $q_-$ characterizing the periods of the profiles $\Phi_+(\mathbf{r})$ and $\Phi_-(\mathbf{r})$ are the locations of the absolute minima of the function $a(q)$ and $b(q)$, respectively, sought within the semiaxis $0 < q^2 < \infty$. The lines (4.2a) and (4.2b) intersect in the point with the coordinates

$$\chi_{AC} = \chi_{AC}^{(I)} = a(q_-), \chi = \chi^{(I)} = (a(q_-) + b(q_+))/4. \tag{4.3}$$

As is seen from Figure 4a, the values of the reduced squared critical wave numbers $q_+$ and $q_-$ for both the linear and miktoarm ABC copolymers are rather different. Thus, the ternary ABC block co-polymers having the χ-parameters in the region *iv*) belong to the class of the copolymer systems revealing two-length-scale behavior.[45,51,56,65,66]

A typical separation of the plane $(\tilde{\chi}_{AC}, \tilde{\chi})$ into the regions with different types of the spinodal instability is shown in Figure 4b. The dependences of the coordinates $\chi_{AC}^{(I)}, \chi^{(I)}$ and the ratio $k = \chi_{AC}^{(I)} / \chi^{(I)}$ of *f* are plotted in Figures 4c,d, respectively.

Now, for simple temperature dependences $\chi_{AC}(T) = \Theta_{AC}/(2T)$, $\chi(T) = \Theta/(2T)$ the states of a ternary ABC system with different temperatures are located on the straight line $\tilde{\chi}_{AC} = k\tilde{\chi}$, $k = \Theta_{AC}/\Theta$ in the plane $(\tilde{\chi}_{AC}, \tilde{\chi})$. Thus, as shown in Figure 4a, depending on the value of *k* the system can leave the stability region crossing either the line (4.2a) or (4.2b). Accordingly, it follows from eqs (2.8), (3.19a) and (3.27) that in the first case, which holds, e.g., for *ABA* copolymer ($\chi_{AC} = 0$), the effective cubic vertex reads

$$\gamma(1) = 2^{-3/2}\left(\Gamma_{111}^{(3)} + \Gamma_{222}^{(3)} + 3\Gamma_{112}^{(3)} + 3\Gamma_{221}^{(3)}\right), \tag{4.4a}$$

(here $\Gamma_{ijk}^{(3)}(1) = \Gamma_{ijk}^{(3)}(\mathbf{q}_1, \mathbf{q}_2, \mathbf{q}_3)$, $|\mathbf{q}_i| = q_*$, $\mathbf{q}_1 + \mathbf{q}_2 + \mathbf{q}_3 = 0$). The straightforward calculation shows, as consistent with refs 15,16,19,59, that for symmetric triblock (miktoarm) copolymer $A_{fN}B_{(1-2f)N}A_{fN}$ the only critical point where the cubic vertex (4.3a) vanishes is located at $f = 0.2551$ ( $f = 0.2215$ ).

On the contrary, in the second case the cubic vertex

$$\gamma(1) = 2^{-3/2}\left(\Gamma_{111}^{(3)} - \Gamma_{222}^{(3)} + 3\Gamma_{221}^{(3)} - 3\Gamma_{112}^{(3)}\right). \tag{4.4b}$$

vanishes identically for symmetric copolymer with any composition *f* of the non-selective block.

Thus, the ternary ABC block copolymers belonging to the AC-modulation class are expected to undergo much smoother ODT than those belonging to the AC-modulation class.

So, before building the phase diagrams we are to decide **which choice of the relationship between the χ-parameters is the most appropriate** for the system under consideration, and, accordingly, to which region on the plane $(\chi_{AC}, \chi)$ belong the copolymers under consideration. Let us start with the assumption that the Flory-Huggins parameters satisfy the **Hildebrand approximation**:

$$\chi_{ij} = v(\delta_i - \delta_j)^2 / T, \tag{4.5}$$

where $\delta_i$ is the conventional solubility parameter of the *i*-th component usually supposed to be temperature-independent. It is convenient[61] to take as the only two independent interaction parameters characterizing the ternary systems in the approximation (4.4) the following ones:

$$\chi = \chi_{AC} = v(\delta_A - \delta_C)^2/T, \quad x = (2\delta_B - \delta_A - \delta_C)/(\delta_A - \delta_C),$$
$$\chi_{AB} = \chi(1-x)^2/4; \quad \chi_{BC} = \chi(1+x)^2/4; \quad \chi_{AC} - \chi_{AB} - \chi_{BC} = \chi(1-x^2)/2, \quad (4.6)$$

$\chi_{AC}$ characterizes incompatibility of the side blocks in the ABC triblock copolymer whereas the *selectivity parameter*[61] $x$ describes how much is the middle block B selective with respect to the side blocks.

In the Hildebrand approximation the symmetry assumption $\chi_{AB} = \chi_{BC} = \chi$ holds in two cases:

i) if $\delta_A = \delta_C$, $\chi_{AC} = 0$, $|x| = \infty$, which corresponds to the ABA block copolymer; and

ii) if $\delta_A - \delta_B = \delta_B - \delta_C$, $k = \chi_{AC}/\chi = 4$, $x = 0$, which means that the middle block (for linear) or one of the blocks (for miktoarm) ABC macromolecule) is non-selective with respect to two others. As shown above, in the latter case the cubic term vanishes identically for any composition *f* of the non-selective block. It is this peculiarity of the ternary linear block ABC copolymers with the middle block non-selective as to two other (*x=0*), which provides continuous ODT transition in these systems and, thus, enables to describe the transition close to the ODT within the WS theory in the next section. It is worth to note that case *ii*) holds for the poly(isopren-*b*-styrene-*b*-2-vinylpyridine) triblock copolymers.[35]

The continuous ODT transition in the ternary block copolymers takes place not only for $f_A = f_C$ and *x=0*. The critical lines build via numerical solving the equation $\gamma(1, f_A, f_C) = 0$ are shown for different values of *0<x<1* for linear ABC copolymers in Figure 5a. We represent[60] the block copolymers $A_nB_mC_l$ as a point in the plane of the side blocks compositions ($f_A, f_C$). All these points lay inside of the right triangle *ABC* shown in Figure 5a. Thereby, the sides *AB*, *BC*, and *AC* represent the compositions of the corresponding diblock copolymers, the diagonal *Bb* corresponds to symmetrical copolymers $A_nB_mC_n$, and the dashed lines *cb* and *ab* represent the linear ABC block copolymers in which the composition of the largest side block ($f_A$ and $f_C$, respectively) is equal to 0.5.

Remarkably, the line $f_A = f_C$ is not the only critical line even for *x=0*. Another critical line is the curve *ac*, which is rather flat one and close to the straight line *ac*. For non-zero selectivity parameter $0 < x < 1$ the critical lines consist of two branches, which in the limit $x \to 1$ approach the lines *ab* and B*c*, respectively. The physical meaning of these branches is clear: the branch approaching the straight line *ab* corresponds to the case when the composition of the block C is close to 0.5 and incompatibility of the A and B blocks is much less than that of each of them with C block (we will refer to the branch as the C(AB)), whereas that approaching B*c* does to AB copolymer with a short strongly interacting C block[60] (we will refer to the branch as the AB(C)).

Similarly, the critical lines $\gamma(1, f_A, f_C) = 0$ for miktoarm ABC copolymers are plotted in Figure 5b, but here we label the most incompatible blocks as the A and C blocks. It is seen that for these systems the critical lines also consist of two branches, which in the limit $x \to 1$ approach the lines *ab* (the C(AB) branch) and B*c* (the AB(C) branch), respectively. But now the limiting branch *ab* (the bold line in Figure 5b), which corresponds to the critical line for the symmetric $A_nB_mB_l$ graft copolymer, is not the straight line since such a copolymer can not be mapped onto the linear AB block copolymer.

The behavior of the critical lines for the linear ABC copolymers with *x>1* also could be analyzed in similar way. However, it is somewhat more complicated and we address it elsewhere.

### 5. The phase diagrams.

Now we apply the procedure described in sections 3,4 to analyze the typical features of the phase behavior of the ternary ABC block copolymers.

**The advantages and limits of the 2$^{nd}$ shell harmonics approximation.** To begin with, we present in Figure 6 the phase diagram of the molten diblock copolymer, which could be considered also as that of molten ABC block copolymer with $f_B = 0$, calculated within our procedure. Remember that it differs from that of Leibler[2] in taking into account the 2$^{nd}$ shell harmonics contribution into the 4$^{th}$ vertex of the effective Landau Hamiltonian within its accuracy only as described in concluding sub-section of section 3. Therefore, comparing our phase diagram with those of Leibler[2] and Matsen and Schick[24] one can estimate both the advantages and deficiencies of our approximation.

As is seen in Figure 6b, as far as the conventional phases is concerned our phase diagram would almost coincide with that of Leibler[2] precisely approaching the latter in the vicinity of the critical point. The only difference would be some broadening of the BCC phase stability region (basically at cost of the HEX phase) with increase of the diblock copolymer asymmetry. The situation changes drastically as soon as we include into the list of competing phases those of the **G family,** which were *not* taken into account in the original paper.[2] All three phases of the family described in section 2 become stable when the asymmetry $|f - f_c|$ increases *and* the 2$^{nd}$ shell harmonics effect is taken into account. It is worth to notice that when we calculate the free energy of the ordered phases within the Leibler 1$^{st}$ harmonics approximation the phases of the **G** family turn out to be metastable only. Therewith, in our approximation the triple point LAM-HEX-G is located at *f=0.462, $\tilde{\chi} = 10.88$*, which is rather close to the result *f=0.452, $\tilde{\chi} = 11.14$* obtained by Matsen and Schick[24] within the SCFT using much more harmonics. Comparing the presented numerical results for the triple point we conclude that our 2$^{nd}$ shell harmonics approximation somewhat *overestimates* the effect of the higher harmonics but, nevertheless, is in a reasonable agreement with the SCFT results obtained using the whole series of the higher harmonics[24].

Even more striking indication of such an overestimation is the sharp falling phase transition line DIS-BCC$_3$ below the spinodal, $\tilde{\chi}_{\text{DIS-BCC}_3}(f) \to -\infty$ when $f \to f_{\text{di}} = 0.4183$. At the first sight this result could seem meaningless. But, in fact, its physical meaning is rather clear. It could be understood by analogy with that of the spinodal of block copolymers with respect to microphase separation. The latter was defined[1-3] as the line (surface) where the minimal eigenvalue of the inverse matrix of the correlation functions appearing in the quadratic term of the free energy (3.7) vanishes. Accordingly, the correlation functions calculated within the random phase approximation (RPA) become divergent here and, thus, the uniform state of the systems becomes absolutely unstable. However, taking into account the fluctuation corrections[26-31] shows that the correlation functions stay finite and the uniform state stays stable (at least metastable) even beyond the RPA spinodal. Never-theless, the latter stays a useful notion when understood as a crossover line between the regions with different temperature scaling of the correlation radius and the exact border of the region where the RPA is applicable at least qualitatively.

Quite similarly, it follows from eq (2.21) that the sharp falling down of the phase transition line DIS-BCC$_3$ when $f \to f_{di}$ is determined by the fact that the minimal quadric vertex $\beta_{\overline{\mathfrak{R}}} = \min \beta_{\mathfrak{R}}$ changes the sign in the point $f = f_d$ due to the 2$^{nd}$ shell harmonics renormalization of the vertex and $\overline{\mathfrak{R}} = \text{BCC}_3$ for molten diblock copolymer. So, $\beta_{\text{BCC}_3}(f) < 0$ for $f < f_{di}$ and, therefore, the expansion of the Landau Hamiltonian up to the 4$^{th}$ term only in powers of the order parameter $\Psi$ becomes inapplicable. As in the spinodal case, the unphysical divergence of the leading term is to be removed by including into the expansion the terms of the higher order than that causing the divergence. In our case it means to take into account *at least* the terms of the 5$^{th}$ and 6$^{th}$ powers in $\Psi$ as well as the 3$^{rd}$ shell harmonics contributions. The corresponding modification, obviously, goes far beyond the scope of the present paper, but it is expected to smooth (not eliminate!) the sharp phase transition line DIS-BCC$_3$ (in general, DIS-$\overline{\mathfrak{R}}$) shown in Figure 6. Further we refer to the line $\beta_{\overline{\mathfrak{R}}} = 0$ as the WS border line since beyond the line the higher harmonics effect becomes so important that the system could not be described properly even within the 2$^{nd}$ shell harmonics approximation of the WS theory.

Two more interesting features of the modified WS phase diagram shown in Figure 6 are the phase transition lines G – BCC$_3$ at *f=0.4343,* which is the root of eq (2.21), and G – G$_2$ situated at relatively high values of $\tilde{\chi}$. It is important to stress that the WS theory can not claim responsibility for prediction of precise location of both these phase transition lines since they lay too far from the critical point. Moreover, the stability of the phases BCC$_3$ and G$_2$ could be only an artifact of the WST extrapolation beyond its validity region. Nevertheless, these phase transition lines are interesting as indications of the fact that it is only a moderate development of the 2$^{nd}$ shell harmonics, which increases the stability of the double gyroid phase G, whereas the further increase of their amplitude (and, thus, degree of segregation) results in increase of stability of other cubic phases (in our case, $BCC_3$ and $G_2$) at cost of the G phase.

Summarizing, the modified WS theory we presented in the paper provides a rather reasonable accuracy in locating the triple point HEX-G-LAM and interesting (much less reliable, though) hints as to stability of some other non-conventional cubic phases. In any case, the presented phase diagram of the molten diblock copolymers establishes a good baseline for further discussion.

**The linear ternary ABC block copolymers.** The general peculiarities of the ternary ABC block copolymer phase behavior could be well seen already for the middle non-selective block (the block interaction is assumed to be described by the Hildebrand approximation (4.6) with $x=0$), which is the case for molten poly(isopren-$b$-styrene-$b$-2-vinylpyridine) triblock copolymers.[35] These peculiarities are presented in Figure 7 and Figure 8.

It is seen from Figure 8 that the evolution of the phase diagrams with increase of the composition of the middle non-selective block $f_B$ is many-sided. First, the width of the composition interval between the WS borders increases with $f_B$ (see Figure 8 a-d) until the WS theory becomes applicable for the whole range of the composition $f_A + f_C = 1 - f_B$ at $f_B = f_{di}$. The location of the triple point HEX-G-LAM shifts towards the critical point and reaches the latter at $f_B = 0.407$ (see Figure 8c). However, the width of the temperature interval between the spinodal and the 1$^{st}$ order phase transition into the lamellar phase and thus, the width of the G phase stability decreases with increase of $f_B$ until all the phase transition lines practically merge at $f_B = f_{crit}^{lin} = 0.492$, which corresponds to the point where both critical lines shown in Figure 7 intersect. So, in this region the double gyroid phase G is hardly observable.

When $f_B$ increases further, the region of the G phase stability becomes rather pronounced (Figure 8e) but starting since $f_B = 1 - f_{di}$ the composition interval within the WS borders decreases again (Figure 8f) until it shrinks to zero for $f_B \to 1$. The most interesting feature of the further phase diagram evolution is replacing of the lamellar phase as the most low-temperature stable phase by the consecutively changing non-conventional cubic phases. These phases are $BCC_2$ in the interval $0.6694 < f_B < 0.7181$ (Figure 8g), FCC in the interval $0.7181 < f_B < 0.7607$ (Figure 8h), SC in the interval $0.7607 < f_B < 0.9663$ (Figure 8, i-k) and BCC$_3$ for $f_B$>0.9663 (Figure 8l). The presented collection of the phase diagrams clearly shows that before to become stable in the critical point with increase of $f_B$ the non-conventional phases first do it far away from the critical point. Thus, one can conclude that even though the phase diagrams are obtained within the WS theory with due regard for the 2$^{nd}$ shell harmonics only, they could be qualitatively reasonable even beyond our approximation.

**The miktoarm ternary ABC block copolymers.** To discuss the general peculiarities of the miktoarm ABC block copolymer phase behavior we use as an example the case when one of the blocks (for definiteness, block B) is non-selective with respect to two other (the block interaction is assumed to be described by the Hildebrand approximation (4.6) with *x=0*). These peculiarities are presented in Figure 9 and Figure 10.

The evolution of the phase diagrams with increase of the non-selective block composition $f_B$ is similar to that for linear ABC block copolymers until $f_B$ stays small (Figure 10, a,b): the composition width of the WS theory applicability increases and the temperature span between the DIS and LAM phases decreases. However, unlike the linear case, the phase transition lines for miktoarm case do not merge even along the line $f_B = f_{crit}^{miktoarm} = 0.3035$ tangent to the one of the critical line branches at the point where the branches intersect (Figure 10, b), which is due to much higher curvature of this branch as compared to the linear case. The further increase of $f_B$ results in consecutive change of the phase diagrams topology when $f_B$ passes the characteristic values $f_{LAM-HEX} = 0.3141$, $f_{HEX-BCC} = 0.3279$, $f_{BCC3} = 0.3757$, $f_{BCC3}^{up} = 0.3806$, $f_{tri} = 0.3815$, $f_{BCC3}^{down} = 0.3961$, $f_{di} = 0.4183$ and $f_{strong}^{mikto} = 0.4339$. First, the second critical point with $f_A \neq f_C$ appears in the interval $f_{crit}^{miktoarm} \leq f_B \leq 0.5$ (Figure 10, c-f). Next, for $f_B > f_{LAM-HEX}$ the difference $\beta_{HEX} - \beta_{LAM}$ becomes negative in the critical point $f_A = f_C = f_B/2$ and, thus, the lamellar phase becomes here less thermodynamically advantageous than the HEX phase (Figure 10c). Then the difference $\beta_{BCC} - \beta_{HEX}$ becomes negative in the critical point $f_A = f_C = f_B/2$ for $f_B > f_{HEX-BCC}$ and, thus, the HEX phase becomes here less thermodynamically advantageous than the BCC phase (Figure 10d). With further increase of $f_B$ the phase diagrams undergo rather interesting transformations between the critical points. At $f_B = f_{BCC3}$ the $BCC_3$ phase becomes stable within a small closed loop (Figure 10e). When $f_B$ increases the loop inflates and stretches upwards (Figure 10f) and then (for $f_B > f_{BCC3}^{up}$) becomes the infinite corridor separating two regions of the LAM phase stability (Figure 10g).

For $f_B > f_{tri}$ the forth vertex $\beta_{BCC}$ itself becomes negative along the diagonal $f_A = f_C$, where the odd terms in the expansion of the total free energy vanish identically due to the system symmetry. It is well known from the theory of phase transitions that the point $\beta = 0$ called tricritical is the point where the 2$^{nd}$ order phase transition line is terminated and the 1$^{st}$ order phase transition line starts. One can conclude that in our case the 2$^{nd}$ order DIS-LAM phase transition line converts into the 1$^{st}$ order DIS-BCC phase transition line in the point $f_A = f_C = f_{tri}/2$. As discussed above the WS theory becomes inapplicable for $f_B > f_{tri}$ in a region of the most incompatible blocks compositions $f_A, f_C$ around the diagonal $f_A = f_C$ (see Figure 9 and Figure 10h) where the minimal forth vertex $\beta_{\overline{\mathfrak{R}}} < 0$ and $\overline{\mathfrak{R}} = BCC$. With further increase of $f_B$ the corridor of the $BCC_3$ phase is both broadening and stretching downwards replacing the LAM phase (Figure 10i) and HEX as well as the BCC phase below the spinodal (Figure 10j).

For $f_B > f_{BCC3}^{down}$ another region in the plane $(f_A, f_C)$ appears where $\beta_{\overline{\mathfrak{R}}} < 0$ but here $\overline{\mathfrak{R}} = BCC_3$. So, for $f_B > f_{BCC3}^{down}$ the WS theory still works only somewhat away from the regions marked by the sharp falling down phase transition lines DIS-BCC$_3$ and DIS-BCC (Figure 10k). Then only the part of the phase diagram around the asymmetric critical point latter stays to be treatable within the WS theory (Figure 10l). At last, for $f_B > 1 - f_{di}$ the order-disorder and order-order transitions in miktoarm ABC block copolymers could not be described without using a strong or intermediate segregation theory at all.

To compare the phase behavior of the linear and miktoarm ABC block copolymers along the very critical line $f_A = f_C = f_B/2$ both with each other and the model system analyzed in section 2 we build the reduced phase diagrams in the plane (the composition of the non-selective block $f_B$, the re-duced temperature $\tau_{eff}$), where $\tau_{eff} = \lim_{\sigma \to 0.5} 4\beta(f_B)(\tilde{\chi}_{spin} - \tilde{\chi})/\gamma^2(f_A, f_C)$ and $\gamma(1, f_A, f_C)$ is the value of the effective cubic vertex (4.4b).

It is seen from Figures 8 and 11a that the phase behavior of the linear ternary ABC block copolymers with the non-selective middle block and equal side ones is rather similar to that of the model system whose angle dependence is given by the phenomenological law (2.27) and the reduced phase diagram presented in Figure 3, the middle block composition $f_B$ playing the role of the angle dependence strength $\delta$. Some difference between the reduced phase diagrams shown in Figures 3 and 11a is, obviously, due to the fact that the real angle dependence of the forth vertex for the linear ternary ABC block copolymers is somewhat different from the model law (2.27).

On the contrary, the reduced phase diagram of the miktoarm ABC block copolymers shown in Figure 11b is even topologically very different from Figure 3. Taking into account that the phase diagrams of the linear and graft binary triblock copolymers ABA have the same topology both in the original[15,16,19] and modified[45] with due regard for the 2$^{nd}$ shell harmonics effects WS theory, we infer that the architecture influence on the phase behavior of the ternary ABC block copolymers is much stronger than that for binary AB copolymers.

It is useful to supplement the phase diagrams presented in Figures 8, 10 and 11 with the plots of the spinodal values of $Q_*$ and reduced $\chi$-parameter along the diagonal $f_A = f_C = (1-f_B)/2$ shown in Figure 12. As seen from Figure 12, the miktoarm copolymers with one non-selective block are more stable with respect to the ODT than the linear ones (remember that the triblock ABA copolymers are, in contrast, *less* stable with respect to the ODT than the trigraft ones[58]). Another interesting feature is that the characteristic scale $Q_*$ for the molten ternary linear ABC block copolymer with the middle non-selective block depends on its composition only rather slightly.

**Which choice of the values of $\chi$-parameters is correct?** The Hildebrand approximation (4.4) we used in the preceding section to calculate the phase diagrams of the ternary linear and miktoarm ABC block copolymers is, of course, not the only choice of the relationship between the $\chi$-parameters. Some authors (see, e.g., refs 43, 59, 67) assume that any values of the parameters are plausible given the ratios $\chi_{AB}/\chi_{AC}, \chi_{BC}/\chi_{AC}$ stay constant. In particular, Matsen[43] studied in detail the molten symmetric linear $A_fB_{1-2f}C_f$ copolymers with $\chi_{AB} = \chi_{BC} = \chi_{AC} = \chi$, for which case our WS theory results in an interesting anomaly. Namely, such copolymers belong to the B- and AC-modulation classes for $f<f_1=0.245$ and $f>f_1$, respectively, the spinodal value of χ-parameter for the separating composition being $\chi(f_1)$=24.11. Obviously, for $f$ close to $f_1$ both fields $\Phi_+, \Phi_-$ are strongly fluctuating and, therefore, the conventional one-field WS theory (even modified as shown in section 3) is not applicable. Remarkably, Matsen[43] found for this case a triple point at *f=0.248, χ=20.38,* which is rather close to our data.

In fact, Matsen tried to explore even a broader region of the values of χ-parameters and presented one section ($\chi N = 50$) of the whole phase diagram in the plane $(f, k = \chi_{AC}/\chi)$ (see Figure 8 in ref 43). But he has explored the region *k<2* only, which corresponds to the B-modulation region only (see Figures 4 b,d and discussion in section 4). The region *k>2* (AC-modulation region) and, in particular, the case *k=4*, which corresponds to the Hildebrand approximation, were skipped in ref 43. In other words, the results[43] would describe the phase behavior of those block copolymers, which would strongly deviate from the Hildebrand approximation and, thus, belong to a completely different class of universality.

The usual justification to assume such a strong deviation from the Hildebrand approximation is to say that it could be caused by some specific interactions, e.g., reversible hydrogen bonds etc. This argument is correct but incomplete. The specific interactions do modify the values of χ-parameters,[68] i.e. of the 2nd term (3.3) of the free energy expansion (3.6) but they modify also the 3rd and 4th terms (3.4) of the expansion, which would renormalize accordingly the basic equations for self-consistent fields used in the SCFT. In particular, the reversible bond formation[69-71] in associating solvents[72] and homopolymer solutions[73] are known to affect considerably the location of the critical points of these systems.

Summarizing, we believe that the Hildebrand approximation is the most suitable one to study the phase behavior of the systems without any specific interactions. Strong deviations from the Hildebrand approximation may be, indeed, observed in the systems with specific interactions but the effect of the latter can not be described properly by a modification of the χ-parameters only. In particular, we address the thermoreversible association effect on the ODT in block copolymers elsewhere.

### 6. The non-conventional morphologies and small-angle scattering.

One of the ways to identify the non-conventional morphologies experimentally is TEM. K.B. Zeldovich elaborated a program to simulate the TEM images for the LAM, HEX, BCC, $BCC_2$, $BCC_3$, G and $G_2$ morphologies within the weak segregation regime and presented their preliminary analysis.[74] Here we discuss which peculiarities of the SAXS or SANS patterns provide distinction between the different non-conventional cubic phases whose dominant harmonics differ in some phase shifts only.

First we address the recent analysis of the scattering patterns in the ordered block copolymer structures by Garstecki and Holyst.[75,76] They simulated the structures by various triply periodic minimal surfaces (TPMS) and calculated the corresponding scattering patterns using the general expression for the scattering intensity in the ordered structures[77]

$$I(\mathbf{q}) = C \overline{\Phi(\mathbf{q})\Phi(-\mathbf{q})} \approx C \overline{\Phi}(\mathbf{q})\overline{\Phi}(-\mathbf{q}) \qquad (6.1)$$

where $C$ is a factor, which depends on the scattering device in use and does not depend on the scattering vector $\mathbf{q}$ ($q = |\mathbf{q}| = (4\pi/\lambda)\sin(\theta/2)$, $\lambda$ is the wave length of the scattered waves and $\theta$ is the scattering angle), the bar implies thermodynamic averaging and the second (approximate) equality holds if the fluctuation-caused (heat) contribution into scattering is negligible as compared to that due to thermodynamically stable modulated order parameter profile.

As is commonly known[77] the strong and narrow (Bragg) peaks of the scattering intensity are observed only for those scattering vectors $\mathbf{q}$, which correspond to non-zero amplitudes $\Phi(\mathbf{q})$ of the Fourier transform of the order parameter profile having certain crystal symmetry. (For some scattering devices the bright circles of the radius $q = |\mathbf{q}|$ would be observed). It is the fact that the sets of the Bragg peaks are different for different crystal lattices[48], which stipulates usefulness of the scat-tering methods. The lists of the existing harmonics and their relative amplitudes $\overline{\Phi}(\mathbf{q})$ corresponding to various TPMS formed in block copolymer melts of different architectures was presented in refs 75, 76. The calculated spectra were reported[76] both to agree and disagree with the experimental data depending on the copolymer architecture.

We believe, however, that such an agreement or disagreement is to large degree occasional. First, the real density profiles[43,66,67] never look like a narrow interface between pure domains. Second, the approach advanced in refs 75, 76 takes into account only the packing factors and neglects the energetic ones. Meanwhile, the relative contribution of the heat fluctuation contribution neglected in (6.1) is strongly temperature dependent. Indeed, the average profile contribution (6.1) itself is temperature dependent since the height and the width of the Bragg peaks scale with the size $D$ of the grain in a polycrystal sample as $D^6$ and $D^{-4}$, respectively,[77] and the average grain size obviously depends on the temperature (at least, because of the temperature dependence of the grain interface energy). On the other hand, the fluctuation caused part of scattering strongly depends on the temperature either. In particular, the scattering due to the composition fluctuations coupling with the transversal sound is anomalously large since it is inverse to the square of the shear modulus[77] small in the weak segregated block copolymers.

As a result, the structure and fluctuation contribution into the total scattering are comparable and the scattering curves $I(q)$ observed in block copolymer ordered phases[21,78] are typically rather smooth as compared to the δ-function-like peaks predicted in refs 75, 76 and reveal only one big dominant peak and a few of relatively small and smooth secondary ones. Nevertheless, the common belief we also adopt is that it is the presence or absence of the secondary peaks, which distinguishes between the different morphologies even in the case of "smeared" scattering curves. Therefore, even though the quantitative statistical calculation of the total scattering intensity with due regard for all the relevant factors mentioned above is beyond the scope of this paper, the 2$^{nd}$ shell harmonics analysis presented in section 3 could identify which of these harmonics for the cubic phases with non-zero phase shifts α (see section 2) are strictly zero and thus provide the experimental way to distinguish these non-conventional phases.

Let us first consider the 2$^{nd}$ shell harmonics for the BCC family, which are presented in the table 1.

**Table I. The 2$^{nd}$ shell harmonics for the BCC family.**

| Harmonics | 1 1 0 | 2 0 0 | 2 1 1 | 2 2 0 |
|---|---|---|---|---|
| Formed as |  | $q_3$-$q_{III}$, -$q_2$-$q_{II}$ | -$q_{II}$- $q_{III}$ | -2$q_{III}$ |
| $q^2/q_*^2$ | 1 | 2 | 3 | 4 |
| BCC | ++ | + | + | + |
| BCC$_2$ (single gyroid) | ++ | − | + | + |

The typical harmonics (the first row) are specified by all the ways to compose them of the 1$^{st}$ harmonics and the relative radius of the corresponding coordination sphere (the second and third row, respectively). The dominant (first) and non-zero 2$^{nd}$ shell harmonics are marked by the double and single plus, respectively. The 2$^{nd}$ harmonic of the BCC$_2$ morphology, which vanishes identically, is marked by minus.

As consistent with our discussion in section 3, they could be generated by the cubic coupling term (3.34) resulting in expression (3.35) for the 2$^{nd}$ shell harmonics amplitudes. As shown in the Table1, the only secondary harmonic, which could be formed by two different pairs of the first ones are those with the wave vectors $\mathbf{Q}_\pm^{(x)} = 2^{-1/2} q_* (\pm 2,0,0)$, $\mathbf{Q}_\pm^{(y)} = 2^{-1/2} q_* (0,\pm 2,0)$ and $\mathbf{Q}_\pm^{(z)} = 2^{-1/2} q_* (0,0,\pm 2)$. It is easy to see that the fact results in enhancement of these harmonics for the BCC morphology, for which the factor $D(\mathbf{Q})$ defined by (3.36) equals 2, and their vanishing for the BCC$_2$ morphology, where appearing in (3.36) contributions have equal absolute values and opposite signs (the phase shifted wave vectors $\mathbf{q}_i$, where $i=1,2,3$, enter into the pairs with the opposite direction and, therefore, opposite phases). Thus, the BCC and BCC$_2$ morphologies are distinguishable by the presence or absence of the 2$^{nd}$ harmonics (2,0,0). Moreover, the latter stay zero even if we take into account the coupling term (3.35b) in the full Hamiltonian. Indeed, in this case expression (3.36) for the amplitude of these harmonics takes the form

$$a_\mathbf{q} = -\left[\left(A_0^2/2\right)D(\mathbf{q})\gamma(h) + \left(A_0^3/6\right)E(\mathbf{q})\right]/\left[C(h-1)^2 + \tau\right], \quad (6.2)$$

where, in addition to the designations introduced in section 3,

$$E(\mathbf{q}) = \Gamma_4(\mathbf{q}_1,\mathbf{q}_2,\mathbf{q}_3,-\mathbf{q}) \sum_{\mathbf{q}_i+\mathbf{q}_j+\mathbf{q}_k=-\mathbf{q}} \exp(i(\phi_i + \phi_j + \phi_k)), \quad (6.3)$$

$\Gamma_4(\mathbf{q}_1,\mathbf{q}_2,\mathbf{q}_3,-\mathbf{q})$ is the value of the forth vertex, which is the same for any of four triples $\{\mathbf{q}_i,\mathbf{q}_j,\mathbf{q}_k\}$ summing up to the second harmonic under consideration:

$$\mathbf{Q}_+^{(x)} = \mathbf{q}_3 - \mathbf{q}_2 - \mathbf{q}_I = \mathbf{q}_I - \mathbf{q}_{II} - \mathbf{q}_{III} = \mathbf{q}_3 + \mathbf{q}_1 - \mathbf{q}_{II} = -\mathbf{q}_1 - \mathbf{q}_2 - \mathbf{q}_{III}.$$

It is seen readily that phases $\Phi_{ijk} = \phi_i + \phi_j + \phi_k$ are zero and $\pi$ for the first two and last two of these triples, respectively, and, therefore, the total sum (6.3) vanishes identically.

Similar canceling of the opposite contributions occurs for the (2,0,0) harmonics for the FCC phase and results in the list of the allowed and prohibited $2^{nd}$ shell harmonics for the G family (see Table 2).

**Table 2. The $2^{nd}$ shell harmonics for the G family.**

| harmonics | Formed of | $\dfrac{q^2}{q_*^2}$ | BCC$_3$ | G (double gyroid) | G$_2$ |
|---|---|---|---|---|---|
| 1 1 0 | $\mathbf{q}_{31} - \mathbf{q}_{02}, \mathbf{q}_{32} - \mathbf{q}_{01}$ | 1/3 | + | - | - |
| 2 0 0 | $\mathbf{q}_{12} + \mathbf{q}_{02}, -\mathbf{q}_{22} - \mathbf{q}_{32}$ $\mathbf{q}_{03} + \mathbf{q}_{13}, -\mathbf{q}_{23} - \mathbf{q}_{33}$ | 2/3 | + | - | - |
| 2 1 1 | $-\mathbf{q}_{11}$ | 1 | ++ | ++ | ++ |
| 2 2 0 | $\mathbf{q}_{03} - \mathbf{q}_{33}$ | 4/3 | + | + | + |
| 3 1 0 | $\mathbf{q}_{31} + \mathbf{q}_{12}, -\mathbf{q}_{01} - \mathbf{q}_{22}$ | 5/3 | + | - | + |
| 3 1 2 | $\mathbf{q}_{31} - \mathbf{q}_{22}$ | 7/3 | + | + | + |
| 4 0 0 | $-\mathbf{q}_{01} - \mathbf{q}_{11}, \mathbf{q}_{21} + \mathbf{q}_{31}$ | 8/3 | + | + | - |
| 3 3 0 | $\mathbf{q}_{21} - \mathbf{q}_{22}, \mathbf{q}_{12} - \mathbf{q}_{11}$ | 3 | + | - | - |
| 4 2 0 | $\mathbf{q}_{21} - \mathbf{q}_{11}$ | 10/3 | + | + | + |
| 3 3 2 | $-\mathbf{q}_{22} - \mathbf{q}_{11}$ | 11/3 | + | + | + |
| 4 2 2 | $-2\mathbf{q}_{11}$ | 4 | + | + | + |

The typical harmonics (the first column) are specified by all the ways to compose them of the $1^{st}$ harmonics and the relative radius of the corresponding coordination sphere (the second and third column, respectively). The dominant (first) and non-zero $2^{nd}$ shell harmonics are marked by double plus and plus, respectively. The $2^{nd}$ shell harmonic of the G and G$_2$ morphology, which vanish identically, are marked by minus.

It follows from the list that:

*i*) The BCC$_3$ scattering pattern is characterized by presence of two secondary Bragg peaks (circles) whose radii $q$ are *less* than the radius $q_*$ of the dominant peak. Similar scattering patterns were also found[75,76] for the ABC linear copolymers basing on the idea of the "P-plumber's nightmare triply periodic minimal surfaces". Therefore, we conclude that it is the type of symmetry rather than the character (weak or strong) of segregation, which determines the scattering pattern.

*ii*) The BCC-BCC$_3$ phase transition could be identified by the sudden appearance of two secondary peaks (circles) whose radii are *less* than the radius of the dominant peak. Another important change observable in monocrystal samples would be transformation of the scattering pattern of six dominant reflections (2.11) into that of 12 dominant reflections (2.17). In the real space this transition would be accompanied by increase of the elementary cell size in $\sqrt{3}$ times.

*iii*) For both the G and G$_2$ phases the secondary peaks whose radii are *less* than the radius of the dominant peak vanish. (Actually, the phase choices (2.19a), (2.19b) are the only compatible with the fact that the peaks are not observed.[44]) Besides, each of the G and G$_2$ phases has two of the 2$^{nd}$ shell harmonics, which vanish identically, one of the prohibited harmonics (3,3,0) being common. Thus, the G→G$_2$ phase transition could be identified by vanishing of the (4,0,0) and appearance of the (3,1,0) harmonics.

*iv*) One more remark concerns the (3,2,1) reflections found[78] in the G phase for PS-PEO diblock copolymer and not observed[21] for PI-PS one. The relative weakness of this reflection as compared to that (4,0,0) observed in both cases[21,78] could be explained by the fact that these harmonics are generated by one and two pairs, respectively, of the first ones (see Table 2). Therefore, the factor (3.37) is doubled and, accordingly, the scattering intensity (6.1) is about four times larger for the latter reflections than for that (3,2,1). For the three-harmonics generation (see eqs (6.2) and (6.3)) this combinatorial factor would be even more pronounced.

To conclude, it is worth to stress that we have not assumed that our morphologies possess certain symmetry to obtain the extinction rules listed in the Tables 1 and 2. Instead, following Leibler[2] we specified the trial functions (2.3a), i.e. some sets of the wave vectors and phase shifts for the dominant harmonics (one coordination sphere) only, and found, for which of the sets the free energy acquires the minimum. The extinction rules were derived afterwards within the WS theory modified with due regard for the 2$^{nd}$ shell harmonics effect. And only finally the crystal symmetry classes corresponding to the sets have been identified by comparing the extinction rules for these classes[48] with the derived ones.

# 7. Discussion.

To better understand the gist of the presented approach let us first remember a popular opinion stated, e.g., in ref 80 as follows: "…within the first harmonics approximation the predictions about ordered structures are limited to classical phases of lamellar, hexagonal and body-centered cubic structures, and consequently the possibility of other structures such as bicontinuous structures, e.g. double gyroid, is excluded…To overcome this limitation the so-called harmonic correction is used by including the higher harmonics". The main concern of the authors guided by this opinion is that an artificial restriction of the number of the harmonics in the free energy calculations could lead to artificial results about the relative stability of the phases. For instance, Huh and Jo[80] demonstrated that a small number of harmonics lead them to believe that a square planar lattice is stable for their system, but including more harmonics completely reversed that result. So, one believes often that as far as any non-conventional phase is concerned, is it a real stable structure or an artificial result of an incorrect harmonics truncation can only be understood by a systematic study using enough number of harmonics. (Actually, the best current implementation of such a thorough harmonics account is the SCFT.[24])

However, the main premise the opinion is based on ("within the first harmonics approximation the predictions about ordered structures are limited to classical phases") is not a universal truth even though it is true for molten diblock copolymers[2] and some other comparatively simple architectures.[15,19,45,80] As established in refs 44-46 and discussed in detail in section 2, the non-conventional structures, including bicontinuous morphologies, *can* be stable even within the first harmonics approximation in case the angle dependence of the 4$^{th}$ vertex is strong enough. On the other hand, as proven in section 3, if the dimensionless amplitudes $\Psi$ of the 1$^{st}$ harmonics are small (which takes place under weak segregation), then the higher harmonics contributions are *small* as compared with those of the 1$^{st}$ harmonics; the higher the harmonic is, the less contribution it provides. Besides, within the first harmonics approximation all phase transition lines merge in the critical point(s), around which segregation is certainly weak. Therefore, if the system is close enough to the critical point(s) then the stability of the non-conventional phases caused by strong angle dependence is per-fectly granted without any extra harmonics. (Obviously, this conclusion is valid only up to the fluctuation corrections[12,26-32] absent in any mean field treatment.)

Thus, to stabilize the non-conventional phases caused by strong angle dependence we do not need the higher harmonics, which was the conventional reason to invoke the latter.[50,63-65,80] Instead, we have two other important reasons. First, we are to take into account the weakly fluctuating order parameter contribution, which is just of the order of $\Psi^4$ (see section 3 and eq (3.25)), to find the actual angle dependence of the renormalized 4$^{th}$ vertex (it is worth to note here that the Leibler[2] phase diagram of molten diblock copolymer can be precisely reproduced after the corresponding reducing the linear ABC system only with this term). Second, we use the 2$^{nd}$ shell harmonics approximation[81] to understand the general trend the higher harmonics change the phase diagrams and to estimate the region where the WST could be used at least qualitatively.

Summarizing, the strategy of the presented approach is follows: *i*) to locate all the critical points in the space of the system parameters; *ii*) to check if the angle dependence of the 4$^{th}$ vertex is strong enough to warrant stability of the non-conventional phases at some of the critical points; *iii*) if yes, then to plot the WS border around the favorable critical points or lines and build the phase diagrams within the WS border. If the calculated regions of the non-conventional phases stability lay within and not too close to the WS border, their actual stability can be considered as well supported (and established close to a critical point). For instance, the triple LAM-HEX-G point for molten diblock copolymers found via our the 2$^{nd}$ shell harmonics approximation in section 5 is in a good agreement with the accurate result by the SCFT[24] even though it is located closer to the WS border than to the critical point.

Following this strategy, we demonstrated in the present paper that the weak segregation theory stays an efficient tool to analyze the phase behavior of the molten ternary ABC block copolymers. When properly modified with due regard for the 2$^{nd}$ shell harmonics contributions, the WST is capable of providing a rather reasonable (as compared to that given by the SCFT[24]) estimation of the triple point HEX-G-LAM location for the molten diblock copolymer. Distinguishing the strongly and weakly fluctuating order parameters for the ternary ABC copolymers provides their classification into B- and AC-modulated structures depending on the relationship between their χ-parameters, the AC-modulated copolymers being shown to undergo typically the ODT into weakly segregated morphologies.

When applied, the numerical calculations within the WST are remarkably fast as compared to those within the SCFT, which enabled us to explore the phase diagrams within a large region of the composition triangles both for linear and miktoarm ABC block copolymers. As the result, a strong composition and architecture dependence of the phase diagram was found. E.g., when increasing the composition of the middle block nonselective with respect to the end ones, the symmetric linear ABC block copolymers reveal a tendency to form consecutively the $BCC_2$ (single non-centrosymmetric gyroid), FCC, SC and $BCC_3$ as the most stable low-temperature phase instead of the LAM phase. In contrast, in the symmetric miktoarm ABC copolymers the LAM phase is replaced consecutively by HEX and BCC, the further increase of the composition of the nonselective block changes the continuous $2^{nd}$ order phase transition to the $1^{st}$ order one into a strongly segregated morphology (presumably, $BCC_3$ or BCC).

Such a strong architecture dependence of the phase behavior for the ternary ABC block copolymers is very different from the situation in the binary AB block copolymers, where the typical phase behavior is qualitatively the same both for linear, graft and star copolymers.[2,15,16,19,45] Since this architecture effect is expected to hold in any ABC (in general, multi-component) block copolymers, which contain more than two types of strongly incompatible blocks, we believe these systems to deserve the special term of *amphiphobic* (in contrast to *amphiphilic*).

As shown in section 5, our modified WST provides also a natural procedure to calculate the boundaries (WS border), beyond which the strongly segregated morphologies only could exist and higher harmonics effect becomes so important that the system could not be described properly even within the $2^{nd}$ shell harmonics approximation of the WST. Both outside the WS border and with the temperature decrease the SCFT should be used instead of the WST but the preliminary WST analysis is expected to provide important hints for the SCFT. In this context the WST prediction of the stable $BCC_3$ phase (*Im3m* whose dominant harmonics belong to the $3^{rd}$ coordination sphere) de-serves a special remark. In most of the presented phase diagrams it is found close to the boundary of the WST validity and, thus, its location (and even the existence itself) is only suggestive. In some other cases (phase diagrams 10$f$ -10$k$ and region 8 in Figure 11a) the BCC$_3$ phase is located around the critical line and, therefore, its existence is certain here. We conclude from topological reasons that the phase could exist in the strong segregation regime either. On the other hand, as mentioned above, the BCC-BCC$_3$ phase transition is nothing but the jump-like (in $3^{1/2}$ times) change of the lattice periodicity when keeping the symmetry. It seems that the SCFT numerical procedure should be specially rebuild to envisage the option of such a jump-like periodicity change.

The WST prediction of the G$_2$ ($I\bar{4}3d$) phase stability is much less reliable since we never found it to be stable close to a critical line yet. Nevertheless, it seems meaningful that the WST predicts the loss of the double gyroid phase G stability with the temperature decrease just about the conditions where the hexagonally perforated lamellar (HPL) phase was observed.[21,79] It would be interesting to study the G$_2$ stability within the SCFT and the relationship between the HPL and G$_2$ phases within the WST.

The last comment is related to the new orthorhombic phases recently observed experimentally[82-85] and found to be stable within the SCFT.[86] In our calculations the phases were not taken into account and we suppose to address the stability of the orthorhombic phases within the WST elsewhere.

**Acknowledgement.** I thank INTAS (Grant INTAS 99-01852) and DFG (SFB 481) for support of the initial part of the research and, especially, Alexander von Humboldt Foundation, which supported finalization of the paper via the Humboldt Research Award. I thank Reimund Stadler[†], Volker Abetz and Kurt Binder for many invaluable discussions and hospitality. I thank also the unknown referees who forced me to complete this full version of the brief communication.[46]

# Figure captions

Figure 1. The planar mapping of the vectors (2.11) (see explanations in the text).

Figure 2. The planar mappings a) of the vectors (2.17) and b) of the edges of the regular icosahedron.
a) the vectors depicted by bold lines have zero phases for all three morphologies G, $G_2$ and $BCC_3$, those depicted by dashed and thin solid lines have phases equal to $\pi$ only for the double gyroid (G) and both for G and $G_2$, respectively (see the definitions (2.19a) and (2.19b);
b) the edges depicted by thin and bold lines correspond to the vectors to be removed and properly rotated to transform the icosahedron into the G cell.

Figure 3. The reduced phase diagram for the model angle dependence (2.27) in the plane (the reduced temperature $\tau_{eff}$ - the angle dependence strength $\delta$). The regions of the stability of the disordered state and body-centered cubic, hexagonal, lamellar, G (double gyroid), $BCC_2$ (single gyroid), face-centered cubic and simple cubic lattices are labeled by the numbers *0, 1, 2, 3, 4, 5, 6* and *7* respectively. The dotted lines $\delta = \delta_2$ and $\delta = \delta_{lim}$ are the asymptotics to the phase transition lines *1-7* and *0-1*, respectively.

Figure 4. The spinodal behavior of the symmetric ternary ABC copolymers:
*a)* the $f_A$-dependences of the reduced squared critical wave numbers $q_-$ (curves *1*) and $q_+$ (curves *2*) for both the linear (solid) and miktoarm (dashed) symmetric ABC copolymers;
*b)* the classification of the spinodal instability regions in the $(\chi, \chi_{AC})$-plane. The solid lines satisfy eqs (4.2a), (4.2b) for $f_A = 0.245$, the numbers *0, 1, 2* and *3* label the stability, AC-, B-modulation and two-length-scale regions, respectively; the dashed lines describe the temperature evolution of the systems with $\chi_{AC}/\chi > k_0$ (*a*), $\chi_{AC}/\chi = k_0$ (*b*) and $\chi_{AC}/\chi < k_0$ (*c*), where $k_0 = \chi_{AC}^{(I)}/\chi^{(I)}$ and $\tilde{\chi}_{AC}^{(I)}, \tilde{\chi}^{(I)}$ are the coordinates of the point of intersection of the solid lines;
*c)* the $f_A$-dependences of the coordinates $\chi_{AC}^{(I)}$ (curves *1*) and $\chi^{(I)}$ (curves *2*) for both the linear (solid) and miktoarm (dashed) symmetric ABC copolymers;
*d)* the $f_A$-dependences of the ratio $k = \chi_{AC}^{(I)}/\chi^{(I)}$ for both the linear (solid) and miktoarm (dashed) symmetric ABC copolymers;

Figure 5. The critical lines for the linear (*a*) and miktoarm (*b*) ternary ABC copolymers in the Hildebrand approximation (4.4) for different values of the selectivity *x*. The symmetric bold lines correspond to non-selective middle block (*x=0*), the critical lines labeled by the numbers *1, 2, 3, 4* and *5* correspond to the values of the selectivity parameter *x=0.01, 0.1, 0.3, 0.5* and *0.8*, respectively.
*a)* the dashed lines *cb* and *ab* are the critical lines for *x=-1* and *x=1*, respectively;
*b)* the bold line *ab* is the critical line for *x=1*.

Figure 6. *a)* the phase diagram of the diblock copolymer with due regard for the 2$^{nd}$ shell harmonics effect. The numbers *0, 1, 2, 3, 4, 8* and *9* label the regions of the stability of the disordered state and body-centered cubic, hexagonal, lamellar, (double) gyroid, $BCC_3$ and $G_2$ lattices, respectively; *b)* comparison with the Leibler[2] phase diagram (shown by the dashed lines).

Figure 7. The map describing the WS theory application to the melts of the linear ABC triblock copolymer with the non-selective middle block (*x=0*) within the Hildebrand approximation (4.6). The critical and WS border lines are shown by bold solid and dashed lines, respectively. The labeled thin solid lines correspond to the compositions of the phase diagrams presented in the accordingly labeled phase diagrams of Figure 6.

Figure 8. The WS phase diagrams of the molten linear ABC block copolymers calculated within the Hildebrand approximation (4.5) in the plane $(\sigma,\tilde{\chi})$, where $\sigma = f_A/(f_A + f_C)$ is the asymmetry parameter, for different values of the non-selective middle block composition: a) $f_B = 0.15$; b) $f_B = 0.3$; c) $f_B = 0.4$; d) $f_B = 0.45$; e) $f_B = 0.55$; f) $f_B = 0.6$; g) $f_B = 0.7$; h) $f_B = 0.75$; i) $f_B = 0.8$; j) $f_B = 0.85$; k) $f_B = 0.94$; l) $f_B = 0.97$. The phases are labeled as in Figures 1 and 4.

Figure 9. The map describing the WS theory application to the melts of the miktoarm ABC triblock copolymer with one non-selective middle block (x=0) within the Hildebrand approximation (4.6). The critical and WS border lines are shown by bold solid and dashed lines, respectively. The labeled thin solid lines correspond to the compositions of the phase diagrams presented in the accordingly labeled phase diagrams of Figure 8.

Figure 10. The WS phase diagrams of the molten miktoarm ABC block copolymers calculated in the Hildebrand approximation (4.5) for different values of the non-selective middle block composition: a) $f_B = 0.15$; b) $f_B = 0.3$; c) $f_B = 0.32$; d) $f_B = 0.35$; e) $f_B = 0.3765$; f) $f_B = 0.379$; g) $f_B = 0.381$; h) $f_B = 0.382$; i $f_B = 0.387$; j) $f_B = 0.39575$; k) $f_B = 0.4$; l) $f_B = 0.5$. The phases designations are the same as in Figures 1 and 4.

Figure 11. The reduced phase diagrams along the critical line $f_A = f_C = f_B/2$ calculated taking into account the actual angle dependence of the ternary ABC block copolymers in the plane (the composition of the non-selective block $f_B$, the reduced temperature $\tau_{\text{eff}}$) for linear (a) and miktoarm (b) ABC block copolymers. The designations of the phases are the same as in Figures 1 and 4.

Figure 12. The dependences of the spinodal values of the reduced $\chi$-parameter (a) and squared critical wave number $Q_*$ (b) on the composition of the non-selective block $f_B$ along the critical line $f_A = f_C = f_B/2$. The solid and solid-dotted lines correspond to the linear and miktoarm ABC block copolymers, respectively. For the miktoarm copolymer the solid and dotted parts of the curves correspond to the regions within and beyond the WS theory validity.

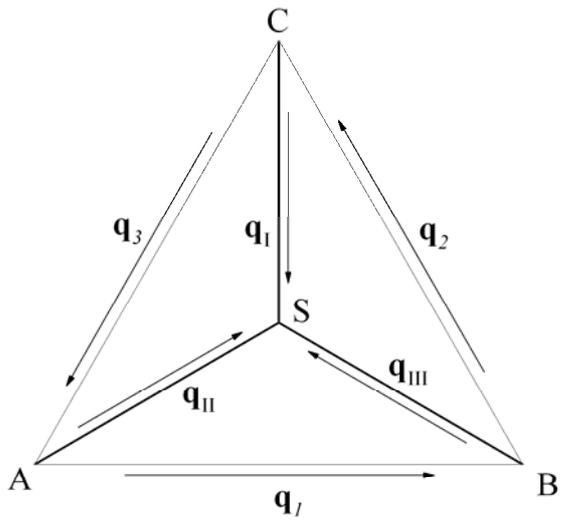

Figure 1.

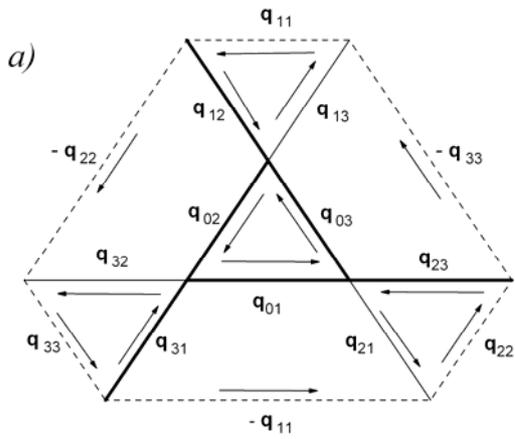 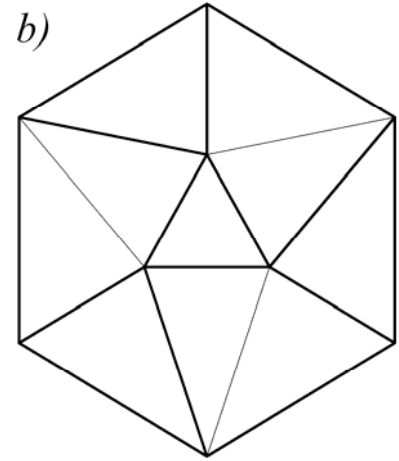

Figure 2.

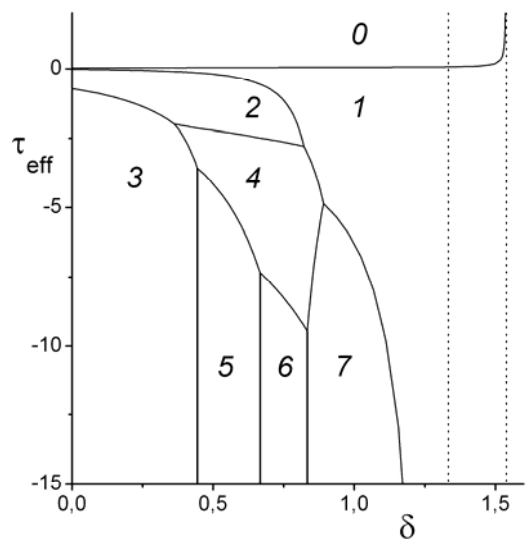

Figure 3.

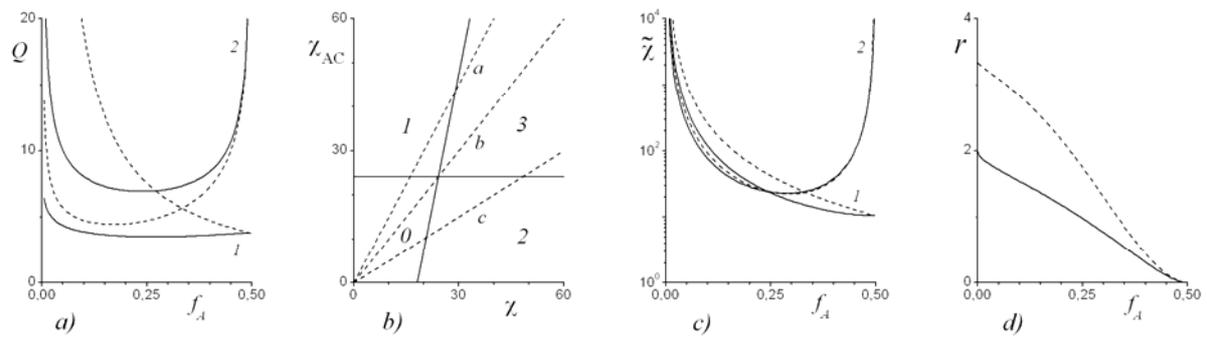

Figure 4.

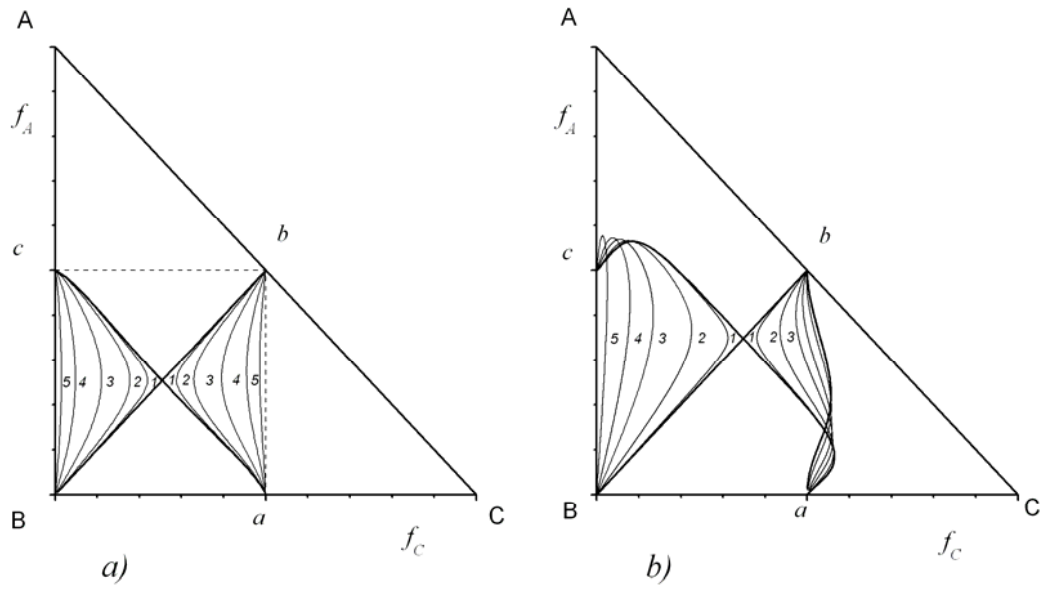

Figure 5.

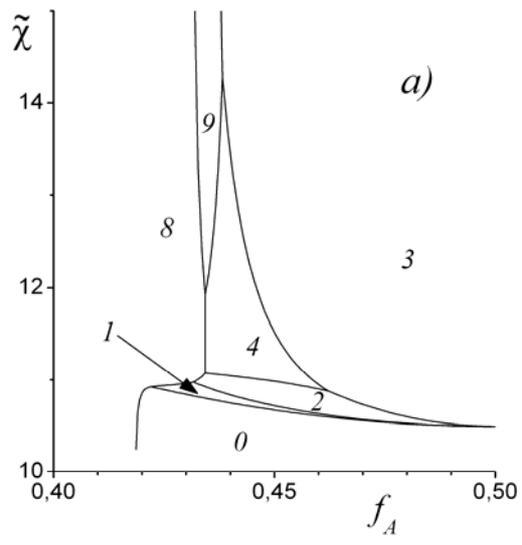 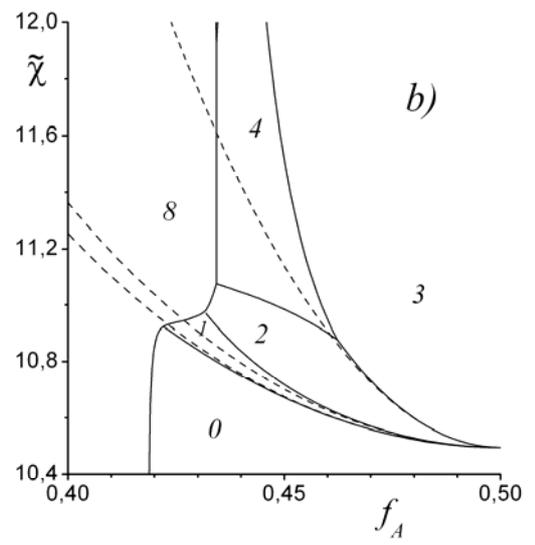

Figure 6.

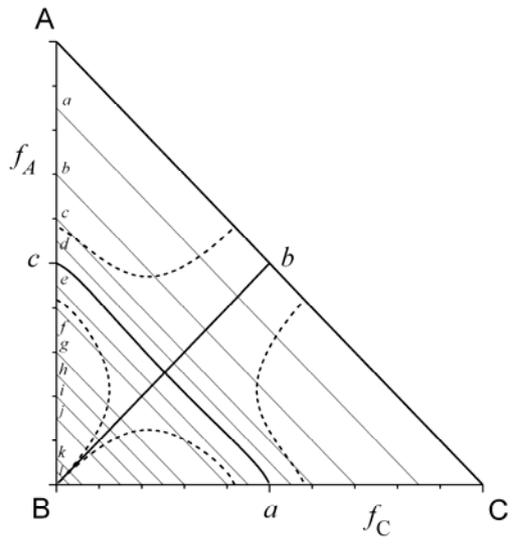

Figure 7.

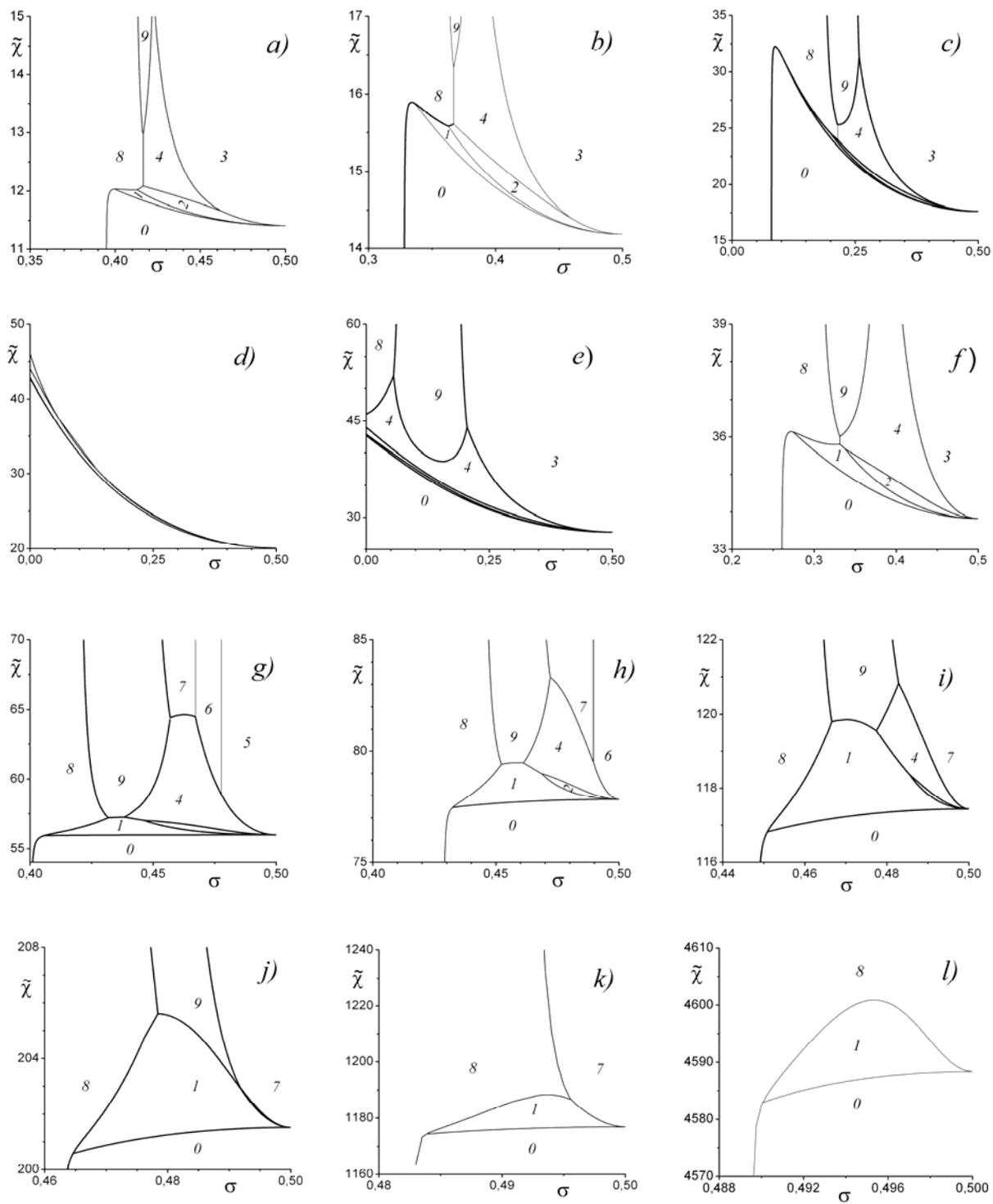

Figure 8.

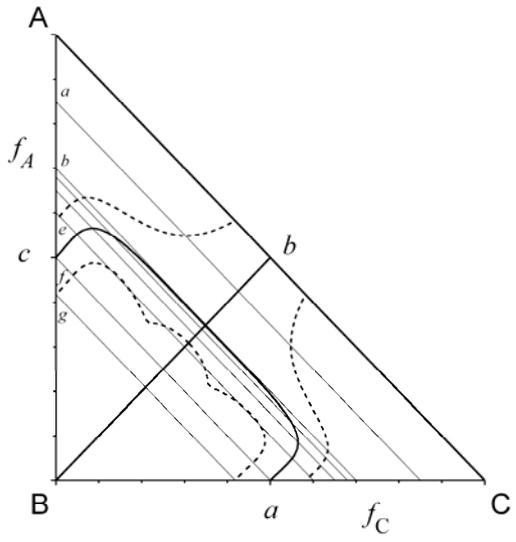

Figure 9.

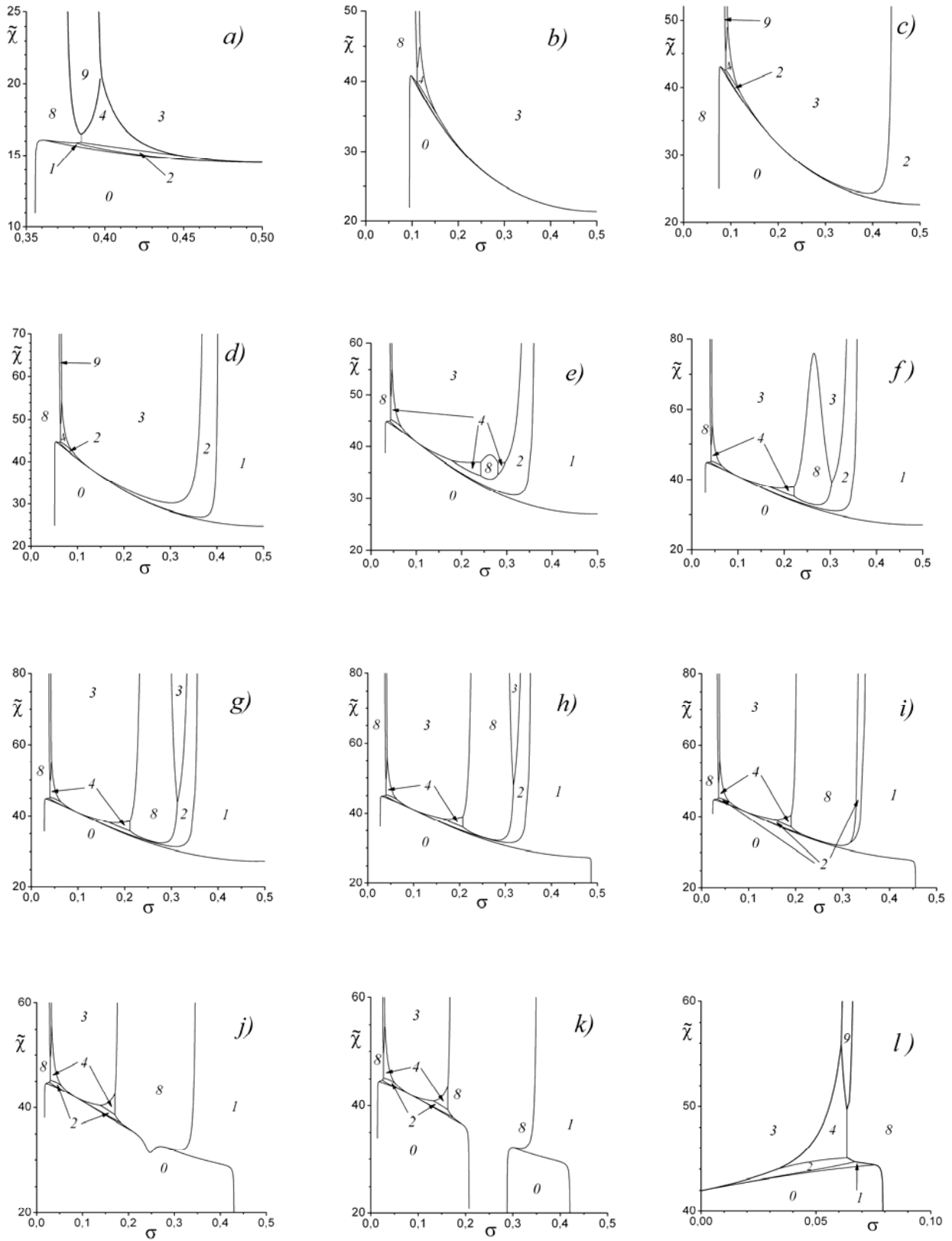

Figure 10.

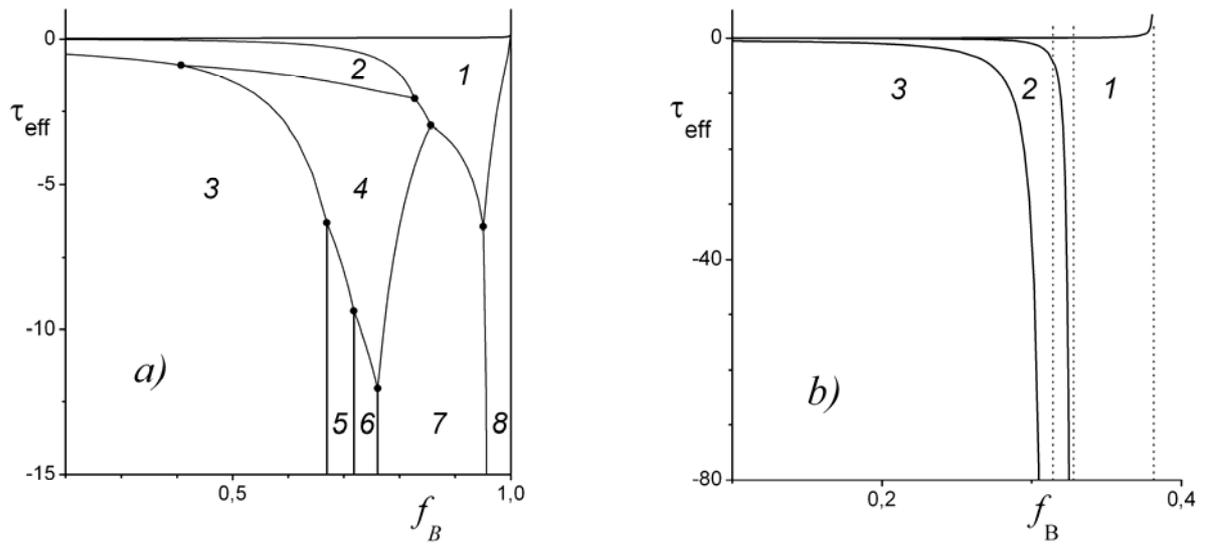

Figure 11.

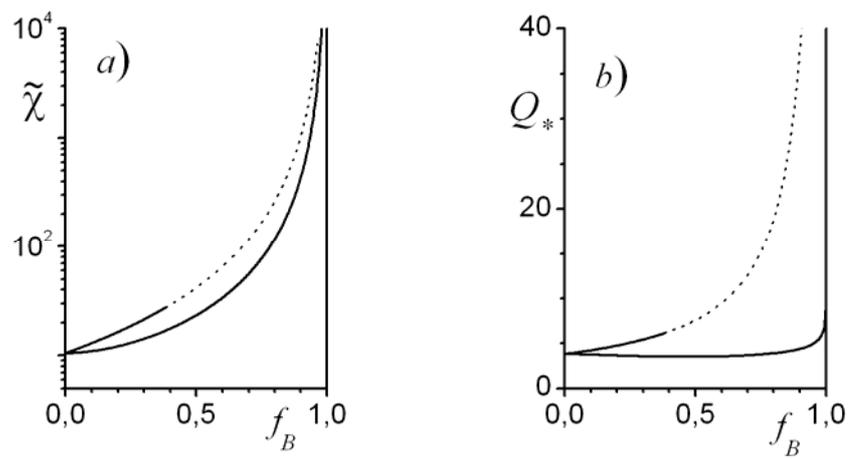

Figure 12.